\documentclass[journal, onecolumn]{IEEEtran}%

\usepackage{amsmath}
\usepackage{amssymb}
\usepackage{graphics,graphicx}
\usepackage{fancyhdr}
\usepackage{xcolor}
\usepackage{todo}
\usepackage{subfigure}
\usepackage{amsfonts}
\usepackage{graphicx}
\usepackage{bbm}
\usepackage[reqno]{empheq}
\usepackage{algorithm}
\usepackage{algorithmic}%
\usepackage{bm}
\usepackage[margin=3cm]{geometry}
\providecommand{\U}[1]{\protect\rule{.1in}{.1in}}
\newtheorem{mytheorem}{Theorem}

\newtheorem{mylemma}[mytheorem]{Lemma}

\newenvironment{myproof}{{\it Proof:$\quad$}}{$\fbox{}$ \\[1em]}

\providecommand{\U}[1]{\protect\rule{.1in}{.1in}}

\newlength{\drop}%

\newcommand*{\titleUL}{
\begingroup
\drop=0.1\textheight
\vspace*{0.5\drop}
\begin{center}
{\LARGE\textsc{$\quad$}}\\[0.5\drop]
\rule{\textwidth}{1pt}\par
\vspace{0.5\baselineskip}
{\huge\bfseries Autonomous Algorithms for Centralized and Distributed Interference Coordination: A Virtual Layer Based Approach 
}\\[0.5\baselineskip]
\rule{\textwidth}{1pt}\par
\vfill
{\Large\textsc{Martin Kasparick$^1$, Gerhard Wunder$^{2}$}}
 \vfill
 $^1$ Fraunhofer Heinrich-Hertz-Institut, Berlin, Germany  (\textit{martin.kasparick@hhi.fraunhofer.de}) \\
 $^2$Technische Unversit{\"a}t Berlin, Germany (also with Fraunhofer Heinrich-Hertz-Institut)  %
\vfill
\today
\vfill
\end{center}
\begin{abstract}
Interference mitigation techniques are essential for improving the performance
of interference limited wireless networks.
In this paper, we introduce novel interference mitigation schemes for wireless
cellular networks with
space division multiple access (SDMA). The schemes are based on a virtual
layer that captures and simplifies the
complicated interference situation in the network and that is used for power
control.
We show how optimization in this virtual layer generates gradually adapting
power control settings that lead to
autonomous interference minimization.
Thereby, the granularity of control ranges from controlling
frequency sub-band power via controlling the power on a per-beam basis,
to a granularity of only enforcing average power constraints per beam. 
In conjunction with suitable short-term scheduling, our algorithms gradually steer the network towards a higher utility.
We use extensive system-level simulations to compare three distributed
algorithms and evaluate their applicability for
different user mobility assumptions. In particular, it turns out that larger
gains can be achieved by imposing average power constraints
and allowing opportunistic scheduling instantaneously, rather than controlling
the power in a strict way. Furthermore, we introduce a centralized algorithm, which
directly solves the underlying optimization and shows
fast convergence, as a performance benchmark for the distributed solutions.
Moreover, we investigate the deviation from global optimality by comparing to a
branch-and-bound-based solution.
\end{abstract}
\begin{center}
\vfill
\textbf{Keywords}: Cellular interference management; SDMA; Distributed algorithms; Autonomous inter-cell coordination;
Power control; Network utility maximization
\vfill
{\itshape Published in EURASIP Journal on Wireless Communications and Networking 2014. Some of the results were presented at conferences \cite{Wunder10, Kasparick2011}.}
\end{center}
\endgroup
}

\begin{document}
    \titleUL
    \newpage

\section{Introduction}

Increasing bandwidth requirements, not least due to the fast growing popularity
of handheld devices with high data rate consumption,
bring cellular networks to the brink of their capacity.
Until recently, an end to the growth of this demand is not yet in sight. In
order to optimally exploit the available bandwidth,
current cellular networks experienced a paradigm shift towards frequency reuse-1.
Consequently, this leads to an increased susceptibility to interference such
that current and future cellular networks are usually interference limited. This
situation is aggravated by a trend towards ever smaller cell sizes.
Especially users at the cell-edge are affected by high inter-cell interference (ICI).
In theory, fully coordinated networks, where neighboring base stations act as a
large distributed antenna array, promise a vast boost in performance \cite{Lee2012, 5gnow2014}.
However, this makes great demands on synchronization and backhaul bandwidth.
In fact, the promised gains from such schemes turn out to be hard to implement
in practice \cite{irmer2011coordinated}.
As a consequence, distributed schemes for interference mitigation,
incorporating
joint scheduling and adaptive power allocation, are of utmost interest. However,
due to mobile users and varying channel conditions, such algorithms have to be
dynamic and able to operate autonomously.
\footnote{Note, that in this paper, we use the term autonomous not in the sense of an autonomous operation of the
different network entities, such as base stations, but to the ability of the network to find a suitable operating
point without the needed of prior planning or human interaction.}
Moreover, future cellular systems,
including pico and femto cells, must be self-organizing to maintain flexibility
and scalability. Therefore, self-optimizing interference coordination schemes
are needed.

In this paper, we introduce such schemes with special focus on cellular space 
division multiple access (SDMA) networks.
All proposed algorithms can be seen as applications of a general radio resource
management framework that configures and optimizes network operation
autonomously and
that allows us to
incorporate service requirements and performance targets.
The framework combines the following three essential ingredients:

\noindent  (i) \textit{Network utility maximization (NUM).}
  By formulating a network utility maximization problem, we steer the system to a desired operating point. 
  This includes fairness goals (like proportional fair and max-min). \\
\noindent (ii) \textit{Virtual model.}
   On top of regular scheduling and resource allocation, we maintain a virtual model of the network that captures
   and simplifies the complicated interference situation in the real world.
   It comprises a static `cooled-down' version of the network based on long-term gains, thus
   suppressing the influence of fast-fading. We use this model to obtain granular power control decisions.
   Thus, it can be seen as a new layer for long-term resource allocation decisions in a cooperative way
   (including corresponding message exchange).\\
\noindent (iii) \textit{Suitable short-term scheduling.}
  Instantaneously, we employ a popular gradient scheduler which is known to asymptotically converge to the solution
  of the underlying utility maximization problem. The scheduler has to take the power constraints into
  account (which can be strict or average constraints) that are obtained in the virtual layer to manage interference.

\begin{figure}[!h]
\centering
%\hrulefill
\vskip-5pt
\includegraphics[width=0.45\linewidth]{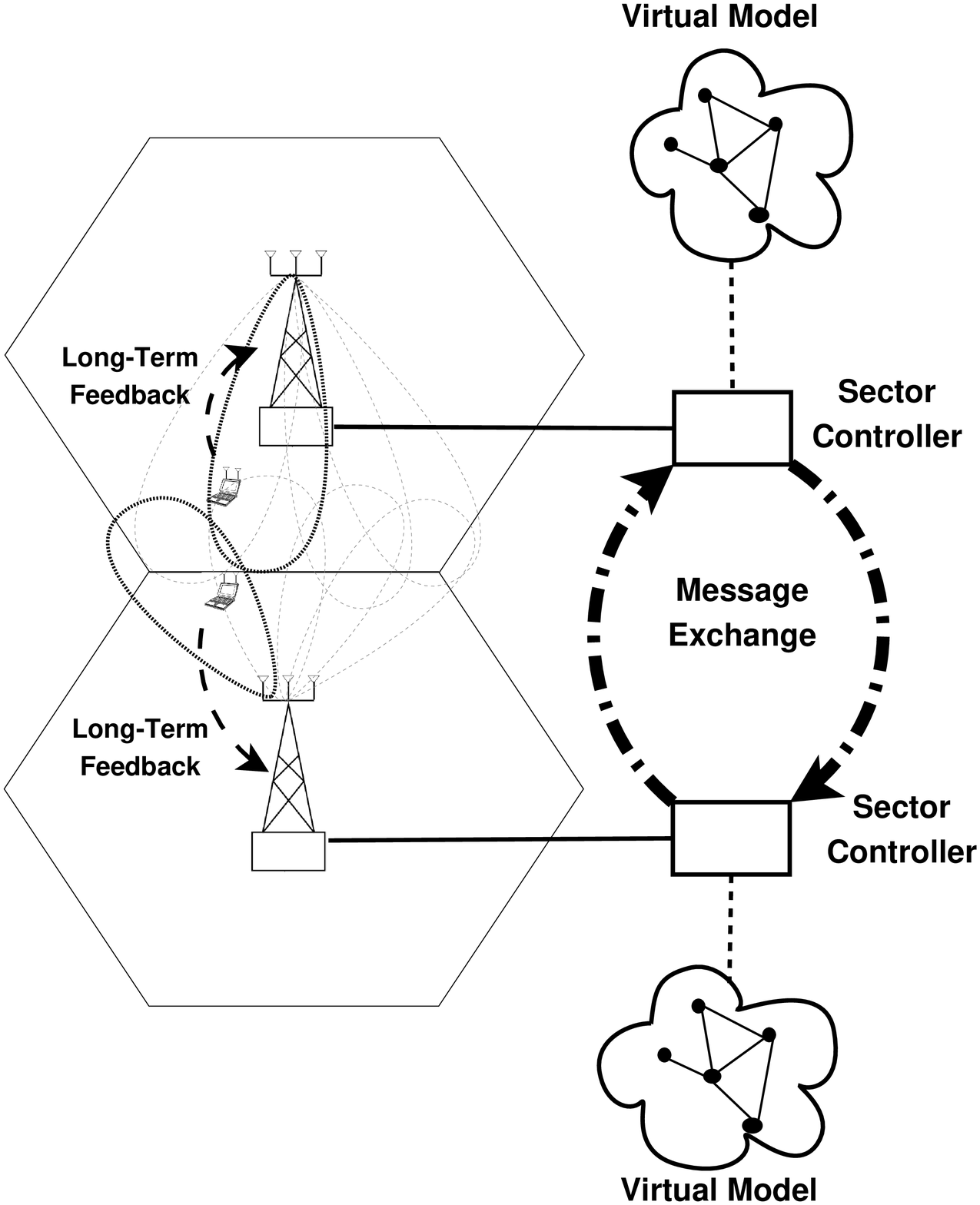}
\caption{General network control approach.
Each sector controller maintains a virtual model of the network based on
long-term feedback. Optimization in this model generates sensitivity
messages which are exchanged among sector controllers and which are used to
adjust power allocations.}
\label{fig1}
\vskip-5pt
%\hrulefill
\end{figure}

To implement this approach, we equip each base station with an additional sector controller that
  requires additional (infrequent) long-term feedback from the mobiles. Note, that this is a practical assumption
  since currently, standardized schemes (such as long-term evolution (LTE)-advanced)
  also consider advanced
  feedback concepts that are not limited to the serving base station \cite{3GPPComp2013}. Moreover, we permit a
  limited message exchange between the sector controllers. Figure~\ref{fig1} depicts the general approach.
 Based upon the long-term feedback, the sector controllers create and update a
`virtual model' of the network. The sector controller is thereby just the entity in each base station that takes
care of all (virtual) resource allocation and scheduling issues.
The main idea of this virtual model is to optimize over `virtual resources', based on
long-term averaged versions of true variables, in order to distributively control a rapidly changing
complex network. Thus, the virtual model captures the complicated interference interdependencies  in the
`real' network and is used for the control and adaption of power allocations. The virtual model allows each sector
to efficiently compute estimates of the gradient of the system utility function with respect to transmit powers of
particular resources in the sector, thus allowing local maximization of the overall utility. More precisely these
estimates are eventually generated by a `virtual scheduling' process. The virtual model and this virtual scheduling
process can be considered as an additional virtual layer for resource allocation. The idea is that if user rates
improve in the virtual layer, they also improve in the real network. Instead of exchanging channel state information
(CSI) with all relevant base stations, only the (gradient) information (called \textit{sensitivities}) that is
obtained in the virtual layer needs to be exchanged. The virtual model, being based on average quantities,
can be further justified since the goal of the algorithm is to adapt the transmit power levels to average interference
levels and not to track fast fading.

As mentioned before, we focus on fixed codebook-based schemes in SDMA networks.
In particular, we assume that each base station maintains a fixed codebook of a
certain size comprising precoding vectors, called `beams'.
These beams can be used to support multiple users on the same time-frequency
resource.
Using fixed codebooks is a practical assumption and allows us to compare our algorithms with
practical schemes used in current cellular systems. 

The optimization within the virtual layer can be organized either in a
centralized or in a distributed manner.
Clearly, a distributed implementation is favorable but in order to accurately
quantify the tradeoffs involved,
we also investigate a centralized solution. This does not only provide a
valuable benchmark for the distributed algorithms,
but may also be a feasible option for small networks, where a central controller
is indeed possible.
Our centralized baseline algorithm is based on an alternating optimization
approach, solving scheduling and power optimization in the
virtual control plane separately.
Since user rates are strongly coupled via transmit powers, we employ a
successive convex approximation technique to tackle the inherent
non-convexity. Due to the non-convex nature of the underlying optimization problem, global
optimality cannot be guaranteed. Therefore, we additionally assess the deviation from global optimality by
comparing our approach to an optimal solution based on branch-and-bound (BNB) in a
simplified setting.

\subsection{Related Work} \label{sec:RelatedWork}
There is a significant amount of research on
interference mitigation in cellular networks \cite{Boudreau2009}, which is often
treated as a specific aspect of self-organization in cellular networks. A
comprehensive survey
on this is provided in \cite{Aliu2012}. A straightforward approach to avoid
interference is to use a frequency reuse factor greater than
one or some fractional frequency reuse scheme \cite{WCM:WCM1088}. Another line
of research targets reuse-1 networks and interference mitigation
by power control and resource allocation.
There have been a variety of suggestions for joint multicell power control and
scheduling in cellular networks such as \cite{Huang2009, Venturino2009}
(see also \cite{gesbert2007} for an overview). Multicell coordination via joint
scheduling, beamforming, and power adaptation is considered in \cite{Yu2011}.
Thereby, fairness requirements (leading to concave utility functions) are
fundamental for current and future cellular standards.
The work \cite{Borst2011} considers joint power allocation and user assignment
to cells in the NUM context, taking into account a mixture of concave and
non-concave utilities.
In \cite{6364444}, a gradient algorithm-based scheme for self-organizing resource
allocation in LTE systems is proposed. However, a multitude of information
has to be exchanged between coordinating sectors.

Although many of the aforementioned references consider distributed schemes,
none treats resource allocation and interference
management in multiuser MIMO systems. By contrast, our framework explicitly aims
to exploit the freedom in terms of resource and power
allocation offered by SDMA. Thereby, our framework builds upon and extends the
framework introduced in \cite{Infocom09, Rengarajan2010} for single-antenna (SISO)
networks.

It is commonly accepted that the underlying optimization problems, which are
non-convex in general, can be solved optimally only for a limited
set of problems and utilities in reasonable time. Existing solutions in cellular
networks often rely on uplink downlink duality \cite{ETT:ETT2554}.
Since they attempt to solve such non-convex problems directly, successive
approximation techniques become an increasingly popular tool to treat this
non-convexity, used for example for power control in DSL
\cite{Papandriopoulos2009} and multihop networks \cite{Papandriopoulos2008}.
Global optimality is often achieved using branch-and-bound-based approaches
\cite{Lin2008,Yu2012} or monotonic optimization \cite{Utschick2012},
however at a high computational complexity.
Other directions include interference pricing \cite{Honig2008,Ahmed2013}
or game theoretic approaches \cite{Sengupta2010}.
Only recently, distributed coordination schemes in cellular networks have gained
increasing popularity \cite{Son10,Infocom09,Venturino2009} due to
the high complexity of centralized approaches.
For example, \cite{6623394} considers the derivation of transmit beamformers, also based on interference prices, for
dense small cell networks.

\subsection{Organization}

The paper is organized as follows. In Section~\ref{sec:SystMod}, we introduce
the considered system model, introduce the notation, and describe the
optimization problem that we address. In Section~\ref{sec:distributedAlgs}, we present a virtual
control layer for solving this problem and introduce three distributed
algorithms that are based on different realizations of this virtual control
plane. In Section~\ref{sec:AO}, we propose an alternative centralized scheme, while in
Section~\ref{sec:BNB}, we present an optimal solution based on branch-and-bound.
In Section~\ref{sec:simulations}, we present system-level simulation results that
evaluate the performance of the distributed algorithms and moreover investigate
simpler scenarios to compare these to the centralized and the optimal
baselines. Eventually, in Section~\ref{sec:Conclusions}, we state the most important
conclusions.

\section{System Model and Notation} \label{sec:SystMod}
We consider the downlink of a cellular OFDMA network, where each cell is
sub-divided into three sectors. In total, we have $M$ sectors $m\in\{1,\ldots,M\}$.
Each sector is served by a base station having $n_\mathrm{T}$ transmit antennas with a
corresponding \textit{sector controller} responsible for user selection and resource
allocation. There are $I$ users randomly distributed in the system, each
equipped with $n_\mathrm{R}$ receive antennas. In the following, we assume $n_\mathrm{R} = 1$.
We assume that each user has pending data at all times. Let $I_m$ be the number of users
associated to sector $m$.

\begin{figure}[h!]
\centering
\includegraphics[width=0.6\linewidth]{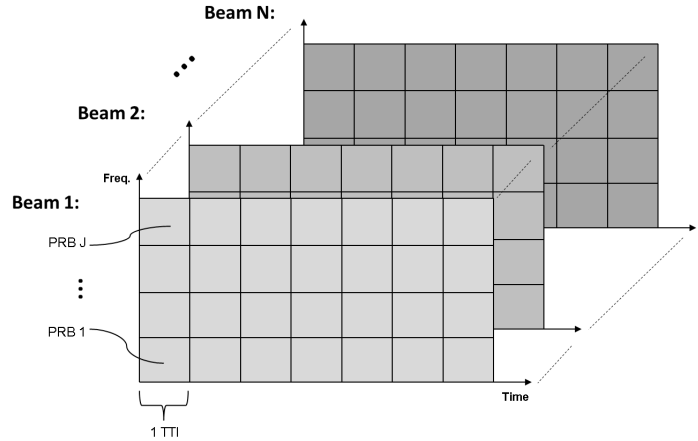}
\caption{Resource grid from the scheduler's perspective. Resource allocation and power control can be performed on a per-PRB-per-beam granularity.}
\label{fig2}
\end{figure}

We assume slotted time with time slots $t=1,2,3,...$, called transmission time
interval (TTI).
In reference to LTE specifications, orthogonal frequency-division multiplexing (OFDM)
sub-carriers are grouped into $J$ sub-bands, called physical resource blocks (PRBs)
\footnote{Each PRB consists of a fixed number of OFDM symbols in time and has a total length of 1 TTI.}.
For the duration of a time slot, the system is in a fixed fading state from a finite set $\mathcal{F}$.
Note, that the assumption of finite fading states is made to foster the analysis.
In our simulations, however, we evaluate the performance using established 3GPP channel models
(cf. Section~\ref{sec:simulations}). Nevertheless, the assumption of a finite set of fading states is often used in
the literature \cite{803503} and can be further justified by the observation that, first, the measurement accuracy of
the user devices is limited and, second, there is only a finite number of modulation and coding schemes that the base
station can choose from. Fading state $l$, in turn, induces a finite set of possible scheduling decisions
$k\in\mathcal{K}\left( l\right)$. We denote by $\pi\left(  l\right)$ the probability of fading state $l$ (with
${\sum_{l}}\pi\left(l\right)  =1$). The sector controllers perform linear precoding, where
precoding vectors for beamforming are taken (as for example in LTE) from a \textit{fixed} $N$-element
codebook $\bm{\mathcal{C}}_{N}:=\left\{  \bm{u}_{1},...,\bm{u}_{N}\right\}$ which is
publicly known. In the following, we identify a beamforming vector ($\bm{u}_b\in\mathbb{C}^{n_{T}}$) by its
index $b\in\left\{1,...,N\right\}$. From the sector controllers' perspective, this makes a beam $b$
on a specific PRB $j$ a possible resource for user selection and power allocation.
This relationship is depicted schematically in Figure~\ref{fig2}.
Let $P_{jb}^m$ be the power assigned to beam $b$ on PRB $j$ by base station $m$.
This value is determined differently in each of the presented approaches under
comparison.
In summary, the (non-trivial) task of each sector controller is to find a
scheduling decision (being an assignment of available resources on PRBs--
to users) and a suitable power allocation, such that a global network utility function is maximized.

Eventually, we have the following additional notational conventions. Let
$\bm{h}_{ij}^{m}\left(  t\right) \in\mathbb{C}^{n_{\mathrm{T}}}$ be the vector of
instantaneous complex
channel gains from base station $m$ to user $i$ on PRB $j$ (we assume
frequency-flat channels within a PRB).
Accordingly, $\left\langle \bm{h}_{ij}^{m}\left(t\right)  ,\bm{u}\right\rangle $
denotes the scalar product of (beamforming) vector $\bm{u}$ and channel
$\bm{h}_{ij}^{m}\left(t\right)$. The noise power at the mobile terminal is
denoted as $\sigma^{2}$. Throughout the paper, we label vectors and matrices
with bold-faced letters. For the ease of reference, the most important notation used throughout this
paper is summarized in Table~\ref{tab1}.

%== Inserted Table 1
%== Table 1 ==
\begin{table}[!h]
\centering
\caption{\bf Important notation}
\label{tab1}
\begin{tabular}[c]{lc}
\hline
\textbf{Notation}                           & \textbf{Definition}\\
\hline
$P_{jb}^m$                                  & Power assigned to beam $b$ on PRB $j$ in sector $m$ \\
$\bar{P}_{jb}^m$                            & Average power constraint (`target' power) of beam $b$ on PRB $j$ in sector $m$ \\
$\bar{P}^m(t)$                              & Current total allocated power in sector $m$ \\
$P_{\text{max}}$                            & Maximum total base station power \\
$c_{jb}(k,\bar{P}_j^m)$                     & Power cost/consumption of beam $b$ on PRB $j$ in sector $m$ \\
$C_{jb}^m(k)$                               & Virtual power cost/consumption of sector $m$ on beam $b$ and PRB $j$\\
$\pi(l)$                                    & Probability of fading state $l$ \\
$\mathcal{K}(l)$                            & Set of possible scheduling decisions given fading state $l$\\
$\bm{h}_{ij}^m(t)$                          & Channel vector from base station $m$ to user $i$ on PRB $j$ at time $t$\\
$\bar{\bm{h}}_{ij}^m$                       & Average channel from base station $m$ to user $i$ on PRB $j$\\
$\bm{\mu}^l_{jb}(k)$                        & Vector of user rates on PRB $j$ and beam $b$ given decision $k$ and
                                              fading state $l$ \\
$\bar{\bm{X}}^m$                            & Vector of average total rates of users in sector $m$\\
$\phi_{jk}^{lm}$                            & Fraction of time that scheduling decision $k$  is chosen on PRB $j$
                                              given fading state $l$ \\
${\phi}_{ijb}^m$                            & Fraction of time that user $i$ in sector $m$ is scheduled using beam $b$
                                              on PRB $j$\\
$G_{ij}^m$                                  & Long-term gain of user $i$ to base station $m$ for his best beam on PRB $j$\\
$G_{ijb}^m$                                 & Long-term gain of user $i$ to base station $m$ for beam $b$ on PRB $j$\\
$R_{ij}^m(k)$                               & Virtual rate of user $i$ on PRB $j$ given decision $k$\\
$\bm{R}_j$                                  & Vector of virtual user rates in PRB $j$ \\
$R_{ijb}^m$                                 & Virtual rate of user $i$ on PRB $j$ and beam $b$ \\
$X_i^m$                                     & Virtual average rate of user $i$ in sector $m$ \\
$\bm{X}(t/{n_v})$                           & Vector of virtual average rates at $t$th virtual scheduling run\\
$F_{ijb}^m$                                 & SINR of user $i$ on beam $b$ and PRB $j$ \\
$D_{jb}^{(\hat{m},m)}$                      & Sensitivity of sector $m$ to a power change of beam $b$ on
                                              PRB $j$ in sector $\hat{m}$ \\
$D_{jb}^m$                                  & Sum sensitivity to a power change of beam $b$ on PRB $j$
                                              in sector $m$\\
$n_j(k)$                                    & Number of beams that are activated on PRB $j$
                                              if decision $k$ is chosen\\
$\lambda_{jb}(t)$                           & Dual parameter: deviation of power on PRB $j$ and beam $b$
                                              from the target value\\
$\alpha_{jb}^m(k)$                          & Scales target beam powers to instantaneous powers `costs'\\
$\alpha_{ijb}^m,\beta_{ijb}^m$              & Approximation constants in  concave lower bound\\
\hline
\end{tabular}
\end{table}

\subsection{Problem Statement} \label{sec:NUM}
The overall goal is to devise autonomous network control schemes which maximize
an overall increasing concave utility function $U$. The utility function is
defined as the sum of sector utility functions $U^m$, which are in turn defined over
\textit{average} user rates $\bar{\bm{X}}^m$. Consequently, the problem to solve for each sector
$m\in\{1,\ldots,M\}$ is given by
%\textcolor{red}{\bf\{AU Query: Please check if all equations were presented correctly. Please note that the standard mathematical notation was used, i.e., italics for single letters denoting mathematical constants, variables, and unknown quantities, and Roman/upright for numerals, operators and punctuations, and commonly defined functions/abbreviations
%(including subscripts). (e.g., $T_{\mathrm{f}} > T_{\mathrm{i}}$, where $T$ is the variable (italicized); `f' and `i' are the labels (upright).\}}
%ANSWER: We checked the equations and made small corrections where necessary, thus the equations are presented correctly. 

\begin{small}
\begin{align}
&  \max_{\bm{\bar{X}}^{m}}\sum_m U^m\left(  \bm{\bar{X}}^{m}\right)
\label{eq:SecUtilBegin}\\
s.t. \;\; & \bm{\bar{X}}^{m}   \leq\sum_{l}\pi\left(  l\right)  \sum_{j}\sum
_{k\in\mathcal{K}\left(  l\right)
}\phi_{jk}^{lm}\sum_{b}\bm{\mu}_{jb}^{l}\left(
k\right) \label{eq:SecUtilRateConstr}\\
&{\displaystyle\sum\limits_{k}}
\phi_{jk}^{lm}=1, \;\text{all}\;l,j,m \label{eq:SecUtilPhiConstr} \\
& 0    \leq\phi_{jk}^{lm}\leq1 \label{eq:SecUtilEnd}
\end{align}
\end{small}

Thereby,
$\bm{\mu}_{jb}^{l}\left(k\right) =
\left(\mu_{ijb}^{l}\left(k\right)\right)_{i\in\{1,\ldots,I_m\}}$ is a vector
comprising elements $\mu_{ijb}^{l}\left(k\right)$,
which represent the rate that user $i$ is assigned on PRB $j$ and beam $b$ when
the system is in fading state $l$ and scheduling decision $k$ is chosen.
They can be zero if the particular resource is not assigned to user $i$ by
decision $k$.
$\phi_{jk}^{lm}$ denotes the fraction of time that scheduling decision $k$ is
chosen on PRB $j$, provided the system is in fading state $l$.

In a nutshell, we are not interested in a specific `snapshot' of the system, but only in ergodic rates.
Therefore, a possible control algorithm should not adapt to a specific system
state but should be able to optimize the system performance \textit{over time}.

Problem (\ref{eq:SecUtilBegin} to \ref{eq:SecUtilEnd}) is solved by applying a
gradient scheduler \cite{Stolyar05} at each time instance. The gradient
scheduler chooses the best scheduling decision $k^\ast$ according to
\begin{small}
\begin{equation}
\label{eq:GradSched}
k^{\ast}\left(  t\right)  \in\arg\max_{k\in\mathcal{K}\left(  l\right)
}\nabla U^{m}\left(  \bm{\bar{X}}^m\left(  t\right)  \right)
{\displaystyle\sum\limits_{b}}
\bm{\mu}_{jb}^{l\left(  t\right)  }\left(  k\right).
\end{equation}
\end{small}

The gradient scheduler tracks average user rates $ \bm{\bar{X}}^m\left(t\right)$ and updates them after each
time slot according to

\begin{small}
\begin{equation}
\bm{\bar{X}}^m\left(  t+1\right)  =\left(  1-\beta\right)  \bm{\bar{X}}^m\left(t\right)
+\beta J
{\displaystyle\sum\limits_{b}}
\bm{\mu}_{jb}^{l\left(  t\right)  }\left(  k^{\ast}\left(  t\right)  \right),
\label{eq:GradSchedVirtRates}
\end{equation}
\end{small}

where the fixed parameter $\beta>0$ determines the size of the averaging window.
The gradient scheduler is known to asymptotically solve the problem (for $\beta\rightarrow 0$) without knowing
the fading distribution. In case of logarithmic utilities, the gradient scheduler becomes the well-known
proportional fair scheduler \footnote{We use $\sum_i\log(\bar{X}_i^m)$ as sector
utility in this paper, leading to a proportional fair operating point.}.

\section{Autonomous Distributed Power Control Algorithms for Interference Mitigation} \label{sec:distributedAlgs}

We now turn to the distributed power control schemes for autonomous interference
management in cellular networks, which are designed to
enable network entities to locally pursue optimization of the global network
utility.
We introduce three basic approaches.
It is important to note that all algorithms are special cases of the well-known
gradient algorithm \cite{Stolyar05,Stolyar2005}, whose convergence behavior has been thoroughly analyzed.
Therefore, we refrain from reproducing this theoretical analysis.  However,
in Section~\ref{sec:simulations}, for illustration purposes, we present numerical results indicating a fast convergence
behavior. The main difference between the algorithms that we propose is the granularity of
power control.

The first algorithm uses an opportunistic scheduler which only adapts the
power per frequency sub-band, which is then distributed equally among activated
beams. We call this \textit{opportunistic algorithm} (OA).
It leaves full choice to the actual scheduler as to which beam
to activate at what time. The scheduler can therefore decide opportunistically
($\Rightarrow$ {opportunistic algorithm}). However, the power budget per PRB which is distributed (equally)
among activated beams is determined by an associated control scheme.

The second algorithm is the \textit{virtual sub-band algorithm} (VSA), which
enforces strict power constraints on each
beam by requiring all beams to be switched on all the time (with power values
given by the associated control). This has the advantage of making the
interference predictable (assuming known power values). However, it leaves only
limited freedom for the actual scheduler, whose task is reduced to user
selection for each beam. Since a beam is always turned on, it can be treated as
an independent resource for scheduling, just like a `virtual' sub-band ($\Rightarrow$ {virtual
sub-band algorithm}).

The third is a hybrid approach, which permits opportunistic scheduling at each
time instance but, in addition, enforces average power constraints per beam.
We call it \textit{cost-based algorithm} (CBA). It leaves more freedom for opportunistic scheduling than the
virtual sub-band algorithm. In contrast to requiring all beams to be used at all
times with strict power values, we only require the target beam power values to be kept \textit{on average}.
Thus, instantaneously, the scheduler is free to make opportunistic decisions based on the current system state.
In order to assure that the average power constraints are kept, we introduce an additional cost term into the utility
maximization and the gradient scheduler ($\Rightarrow$ {cost-based algorithm}).

We focus on the applicability in different fading environments, 
comparing the overall performance with respect to a network-wide utility
function as well as the performance of cell-edge users.
Thereby, we show that although the
algorithms behave differently in different user mobility scenarios, in general,
it is more beneficial to impose average rather than strict power constraints.

 As we demonstrate later, the three algorithms perform differently with
different mobility assumptions on the users.
 A problem that arises with increased mobility is that the (virtual) model lacks
behind the actual network state.
 Especially when the controllers are restricted to a gradual power adaption
process on a per-beam granularity, they might not always be able to fully
exploit multiuser diversity.
 Since the proposed algorithms put different emphasis on opportunistic
scheduling in power adaption and resource allocation
 decisions, they perform differently when facing user mobility and fast
fading.

Let us now turn to the control plane, the virtual layer, of the considered algorithms. They all have
the following general procedure in common.
The goal of the control plane is to obtain estimates of the partial derivatives
of the network utility with respect to the power allocation of
particular resources and to adapt the power control policy accordingly. These
estimates can be seen as estimates of the \textit{sensitivity} of the network
utility to changes of the allocation strategies. Thereby, the allocation can be,
as in the OA of Section~\ref{sec:OA}, the power allocation of a PRB which is then divided
equally among activated beams. Or it can even be, as in the VSA of Section~\ref{sec:VSA}, the
power allocation of an individual beam. Or it can also be, as in the CBA of Section~\ref{sec:CBA},
simply an average power constraint of an individual beam, which does not have to be kept
at every single time instance.

A further similarity between all algorithms is that in addition to ordinary
\textit{short-term CSI}, the mobiles (not necessarily often) report long-term
feedback to their base stations, which is then
used to calculate (virtual) user rates and, accordingly, (virtual) average rates.
These average rates are a good representation of the interference
coupling throughout the network and are used to
calculate the `sensitivities' to power changes on particular resources in
other sectors (and in the own sector). This sensitivity information is compiled
in messages which are exchanged between the sectors\footnote{To reduce messaging overhead,
sector controllers could limit the message exchange to strongest
interferers, e.g., consider only neighboring base stations.}. Upon reception of
the message vectors, the sector controllers are now able to calculate the
desired estimate of the \textit{system} utility's sensitivity to power changes on particular
resources and adjust powers (or power constraints) accordingly.
It is important to always distinguish between actual rates and virtual rates.
The actual average rates are the ones tracked by the ordinary
(proportional fair) scheduler and determine the sector's utility.
The virtual average rates are based on long-term feedback of averaged channel
gains and do not have an immediate physical meaning in the 'real world'.
They are created by a `virtual' scheduler based on `virtual' scheduling decisions.
The set of all these rates forms a `virtual' model of the system which is used to derive power
adaption decisions.

The question remains how the virtual average rates and accordingly the
sensitivities to power changes are calculated. Besides the granularity of
power control, this virtual model, which can be also seen as a virtual layer for
interference mitigation above the actual short-term scheduling, is the main difference between the
investigated algorithms. In the following, we will discuss this in detail.

\subsection{Opportunistic Algorithm (OA)} \label{sec:OA}
OA can be seen as a straightforward extension of the multi-sector gradient (MGR) algorithm in
\cite{Infocom09} to multi-antenna networks. %\textcolor{red}{\bf\{AU Query: For clarity, please provide the expanded form of
%`MGR' since it occurred more than once in the text.\}} 
%ANSWER: The expanded form 'multi-sector gradient' of MGR was added above. 
Although it is designed for multi-antenna networks, it does not
perform power control on a per-beam basis (as opposed to the other two
algorithms) but gives complete freedom to the sector controllers with respect
to the number and choice of beams that are active at every given time instance.
The only value that is controlled is the power budget \textit{per PRB}. Nevertheless,
SDMA is applied where multiple users can be scheduled on the same PRB, however on different beams.
Consequently, the long-term feedback of users comprises a codebook index
(being the maximizing index $b^{\ast}$ in (\ref{eq:GainOA})) as well as a corresponding gain

\begin{equation}
G_{ij}^m=\max_{b}\left\langle \bm{\bar{h}}_{ij}^m,\bm{u}_{b}\right\rangle,
\label{eq:GainOA}
\end{equation}

where $\bm{\bar{h}}_{ij}^{m}$ is the channel from user $i$ to its sector
controller on PRB $j$, averaged to eliminate the influence of fast-fading.

The task of finding scheduling decision $k$ now amounts to determining the best
subset of users to be scheduled on a PRB, subject to the constraint that
each user can only be scheduled exclusively on its reported beam. Let us define
(virtual) user rates (time index omitted),
assuming a user is scheduled on PRB $j$ (otherwise $R_{ij}^m\left(k\right)=0$), given by

\begin{equation}
R_{ij}^m\left(  k\right)  =\rho\left( F_{ij}^{m} \right), \text{with } F_{ij}^{m}:=\frac{G_{ij}^m\bar{P}_{j}^m\left(
k\right)  }{\sigma^{2}+\sum_{m^{\prime}\neq
m}\bar{P}_{j}^{m^{\prime}}{G}_{ij}^{m^{\prime}}}. \label{eq:F_OA}
\end{equation}

In this paper, we use $\rho(x)=\log(1+x)$. Thereby,
$\bar{P}_{j}^m\left(k\right)$ is the current power value of PRB $j$ divided by
the number of users scheduled, thus depending on decision $k$.
Moreover, $\bar{P}_{j}^{m^{\prime}}\tilde{G}_{ij}^{m^{\prime}}$ represents a
long-term estimate of the interference of sector $m^{\prime}$ on PRB $j$ (which
can be measured by the mobiles).

Having defined the virtual user rates $\bm{R}$, virtual \textit{average} user
rates $\bm{X}$ and sensitivities are derived by \textit{virtual scheduling}
(based on a similar procedure in \cite{Infocom09}) as follows.
We run the following steps $n_{v}$ times \textit{per TTI} in each sector $m$ and
for any PRB $j$. Thereby, the parameter $n_{v}$ determines how long
the virtual scheduler runs before accepting the sensitivities. Consequently, a
larger value means more overhead by the virtual layer but
better results.

\begin{list}{\labelitemi}{\leftmargin=1em}
\item We determine the virtual scheduling decision $k^{\ast}$ using a gradient
scheduler according to
\begin{small}
\[
k^{\ast}\in\arg\max_{k}\nabla U^m\left(  \bm{X}^m\left(  \frac{t}{n_{v}}\right)
\right) \bm{R}_{j}\left(  k,\frac{t}{n_v}\right).
\]
\end{small}

\item We update virtual average user rates according to
\begin{small}
\[
\bm{X}\left(  \frac{t+1}{n_{v}}\right)  =\left(  1-\beta_{1}\right)  \bm{X}
\left(  \frac{t}{n_{v}}\right)  +\beta_{1}J \bm{R}_{j}\left(  k^{\ast
},\frac{t}{n_v}\right).
\]
\end{small}

\item We update sensitivities according to
\begin{small}
\begin{equation}
D_{j}^{\left(  \hat{m},m\right)  }\left(  \frac{t+1}{n_{v}}\right)  =
\left(
1-\beta_{2}\right)  D_{j}^{\left(  \hat{m},m\right)  }\left(  \frac{t}{n_{v}
}\right)  +\beta_{2}\sum_{i=1}^{I_m}\frac{\partial U^{m}\left(  X^{m}\right)
}{\partial X_{i}^{m}}\frac{\partial R_{ij}^{m}\left(
k^{\ast},\frac{t}{n_v}\right)
}{\partial\bar{P}_{j}^{\hat{m}}\left(  t\right)  }. \label{eq:OA-SensUpdate}
\end{equation}
\end{small}
\end{list}

Thereby, $\beta_1$ and $\beta_2$ are small averaging parameters.
Using (\ref{eq:F_OA}), the derivates in (\ref{eq:OA-SensUpdate}) are given by
\[
\frac{\partial R_{ij}^{m}}{\partial P_{j}^{\hat{m}}}=\left\{
\begin{array}
[c]{cc}
\rho^{\prime}\left(  F_{ij}^{m}\right)  \frac{F_{ij}^{m}
}{P_{j}^{\hat{m}}} & \hat{m}=m\\
-\rho^{\prime}\left(  F_{ij}^{m}\right)  \frac{\left(
F_{ij
}^{m}\right)
^{2}}{P_{j}^{m}}\frac{G_{ij}^{\hat{m}}}{G_{ij
}^{m}} & \hat{m}\neq m
\end{array}
\right.
\]

Starting with equal power, the adaption of the PRB powers can be summarized as
follows. From time to time, the sensitivities are exchanged and
summed up by each sector controller for each beam and PRB. Since each
$D_{j}^{\left(  \hat{m},m\right)  }$ is an estimation of the sensitivity of
sector $m$'s utility to a
power change in sector $\hat{m}$, the summation gives an estimate of the
\textit{network} utility's sensitivity.
Then, the power is increased on the PRB with the largest positive sum and
decreased on the PRB with the largest negative sum.

\subsection{Virtual Subband Algorithm (VSA)} \label{sec:VSA}
VSA requires long-term feedback that comprises average link gains \textit{per beam} and sector from each mobile.
We define
\begin{equation}
G_{ijb}^m = \overline{\left\vert \left\langle
\bm{h}_{ij}^{m},\bm{u}_{b}\right\rangle\right\vert ^{2}} \label{eq:avgGains}
\end{equation}
to be the average link gain (the bar denotes empirical averaging
over time) of mobile terminal $i$ on beam $b$ and PRB $j$ to sector controller
$m$.
Based on the long-term CSI, the control plane of VSA responsible for determining
the power allocation per beam works as follows.
Given average gains in (\ref{eq:avgGains}), each sector controller calculates
corresponding virtual rates according to
\[
R_{ijb}^{m}=:\rho(F_{ijb}^{m}),\quad\text{with }
\]

\begin{equation}
F_{ijb}^{m}=\frac{G_{ijb}
^{m}P_{jb}^{m}}{\sigma^{2}+\sum_{b^{\prime}\neq b}G_{ijb^{\prime}}
^{m}P_{jb^{\prime}}^{m}+\sum_{\hat{m}\neq m}\sum_{b^{\prime\prime}}P_{jb^{\prime
\prime}}^{\hat{m}}G_{ijb^{\prime\prime}}^{\hat{m}}}. \label{eq:F}
\end{equation}

Note that since all beams are activated at all times, we have an additional
intra-sector interference term (as opposed to OA), since this interference can
no longer be eliminated, e.g., by switching off beams.
Let $X_i^m$ be the virtual average rate of user $i$ in sector $m$ (not
to be confused with actual average rates $\bm{\bar{X}}$ in
(\ref{eq:GradSchedVirtRates})), defined as

\begin{equation}
X_{i}^{m}=\sum_{j=1}^{J}\sum_{b=1}^{N}\tilde{\phi}_{ijb}^{m}R_{ijb}
^{m}.\label{eq:4}
\end{equation}

Here, $\tilde{\phi}^m_{ijb}$ represent \textit{optimal} time fractions of
resource usage for sector $m$. They are determined as a solution to the following optimization problem (for
fixed virtual user rates):

 \begin{align}
 &  f(\bm{R}):=\max_{\phi_{ijb}^{m}}{\sum_{i}U^m\left(  \sum_{j}\sum_{b}{\phi
 _{ijb}^{m}R_{ijb}^{m}}\right)  }\label{eq:VSAoptimization}\\
 s.t.\quad &  \sum_{i}{\phi_{ijb}^{m}=1},\quad0\leq\phi_{ijb}^{m}
 \leq1.\nonumber
 \end{align}

We rely on an explicit solution to (\ref{eq:VSAoptimization}) since we cannot
apply the virtual scheduling from \cite{Infocom09}. This is because
the resources for power control (which are now individual beams) are no longer
orthogonal but cause interference to each other \cite{Wunder10}.
Having the virtual user rates, the sector controllers calculate sensitivities to
power changes on beams for all sectors (including self) and beams, given by

\begin{align}
D_{jb}^{(\hat{m},m)} &=
\sum_i{\frac{\partial U^{m}}{\partial X_{i}^{m}}\frac{\partial
X_{i}^{m}}{\partial P_{jb}^{{\hat{m}}}}} \label{eq:CalcSensitivities} \\ &=
\sum_{i}{\frac{\partial U^{m}}{\partial X_{i}^{m}}
\sum_{b^{\prime}}{\left(\tilde{\phi}^m_{ijb^\prime}\right)^{1-\varepsilon}
}}{{\frac{\partial
R_{ijb^{\prime}}^{m}}{\partial P_{jb}^{\hat{m}}}}}, \nonumber
\end{align}

where

\begin{small}
\[
\frac{\partial R_{ijb^{\prime}}^{m}}{\partial P_{jb}^{\hat{m}}}=\left\{
\begin{array}
[c]{cc}
\rho^{\prime}\left(  F_{ijb^{\prime}}^{m}\right)  \frac{F_{ijb^{\prime}}^{m}
}{P_{jb^{\prime}}^{\hat{m}}} & \hat{m}=m,b^{\prime}=b\\
-\rho^{\prime}\left(  F_{ijb^{\prime}}^{m}\right)  \frac{\left(
F_{ijb^{\prime}
}^{m}\right)  ^{2}}{P_{jb^{\prime}}^{m}}\frac{G_{ijb}^{m}}{G_{ijb^{\prime}
}^{m}} & \hat{m}=m,b^{\prime}\neq b\\
-\rho^{\prime}\left(  F_{ijb^{\prime}}^{m}\right)  \frac{\left(
F_{ijb^{\prime}
}^{m}\right)
^{2}}{P_{jb^{\prime}}^{m}}\frac{G_{ijb}^{\hat{m}}}{G_{ijb^{\prime}
}^{m}} & \hat{m}\neq m
\end{array}
\right.
\]
\end{small}

Note that the small coefficient $\varepsilon > 0$ stems from an application of
Theorem \ref{Thm:2} (which can be found at the end of Section~\ref{sec:CBA}) in order to
ensure the differentiability of problem (\ref{eq:VSAoptimization}).
The such generated sensitivities are exchanged from time to time between all
sector controllers. Thereby, every sector $k$ receives $J\cdot N$ sensitivity
values from all other $(M-1)$ sectors, in addition to the $J\cdot N$ values
from its own sector. Thus, we have

\begin{equation}
\label{eq:CollectSensitivities}
D_{jb}^{m}=\sum_{\hat{m}=1}^{M}D_{jb}^{m,\hat{m}},
\end{equation}

summing up the sensitivities of all sectors (including itself) to a power
change of beam $b$ on PRB $j$ in sector $m$ and which can be either positive
or negative. Note that sector indices $m$ and $\hat{m}$ in the RHS of
(\ref{eq:CollectSensitivities}) are interchanged compared with
(\ref{eq:CalcSensitivities}), since in (\ref{eq:CalcSensitivities}), we are
interested in how the beam in sector $\hat{m}$ interferes with sector $m$, while
in (\ref{eq:CollectSensitivities}), it is of interest how the beam in sector $m$
interferes with (all) sector(s) $\hat{m}$.

Since $D_{jb}^{m,\hat{m}}$ represent estimates of the sector utilities to a
power change on $jb$ in sector $m$, $D_{jb}^{m}$ clearly is an
estimate of the sensitivity of the \textit{system's} utility to a power change
on the respective beam. Depending on the $D_{jb}^{m}$, we can now make a power adjustment
which steers the system operating point towards a greater utility in the
virtual model.

The power adjustment is carried out in steps of $\Delta>0$, which is small and
fixed. Let $\bar{P}^{m}(t)$ denote the current total allocated power in sector
$m$ and $P_{\text{max}}$ the upper bound on the total sector powers. Then, the following
procedure is applied.

\noindent (1) Pick a virtual resource $(jb)_{\ast}$ (if there is one)
such that $D_{(jb)_{\ast}}^{\left(  m\right)  }\left(  t\right)  $ is the smallest
among all virtual resources $jb$ with $D_{jb}^{\left(  m\right)  }\left(t\right)  <0$
and $P_{jb}^{m}\left(  t\right)  >0$. Now, set \[P_{(jb)_{\ast}}^{m}\left(  t+1\right)
= \max\left\{  P_{(jb)_{\ast}} ^{m}\left(  t\right)  -\Delta,0\right\}.
\]

\noindent (2) If $\bar{P}^{m}\left(  t\right)  <P_{\text{max}}$, pick
$(jb)^{\ast}$ (if there is one) such that $D_{(jb)^{\ast}}^{\left(  m\right)  }\left(t\right)$
is the largest among those $jb$ with $D_{jb}^{\left(  m\right)}\left(  t\right)  > 0$.
Set
\[
P_{(jb)^{\ast}}^{m}\left(  t+1\right)  = P_{(jb)^{\ast}}^{m}\left(  t\right)
+\min\left\{  \Delta,P_{\text{max}}-\bar{P}^{m}\left(  t\right)  \right\}.
\]

\noindent (3) {If $\bar{P}^{m}\left(  t\right)  =P_{\text{max}}$ and $\max_{jb}
D_{jb}^{\left(  m\right)  }\left(  t\right)  > 0$, pick a pair $\left((jb)_{\ast},(jb)^{\ast}\right)$
(if there is one) such that $D_{(jb)^{\ast}}^{\left(  m\right)  }\left(  t\right)$ is the largest,
and $D_{(jb)_{\ast}}^{\left(  m\right)  }\left(  t\right)$ is the smallest among those virtual
resources $jb$ with $P_{jb}^{m}\left(  t\right)  > 0$ and
$D_{(jb)_{\ast}}^{\left(  m\right)  }\left(  t\right)  <D_{(jb)^{\ast}}^{\left(  m\right)}\left(  t\right)$.
Set
\begin{align*}
P_{(jb)_{\ast}}^{m}\left(  t+1\right)   &  =\max\left\{  P_{(jb)_{\ast}}
^{m}\left(  t\right)  -\Delta,0\right\}  ,\quad\text{and}\\
P_{(jb)^{\ast}}^{m}\left(  t+1\right)   &  =P_{(jb)^{\ast}}^{m}\left(
t\right)  +\min\left\{  \Delta,P_{(jb)_{\ast}}^{m}\left(  t\right)  \right\}.
\end{align*}
}

By intuition, the algorithm reallocates power to the beams with large positive
utility-sensitivity.
Note that for numerical reasons, it may be necessary to specify a certain
minimum power per beam $P^{\text{min}}_b$ instead of allowing beam powers to
be reduced to zero. In this case, the changes to the algorithmic notation above
are straight forward, so we do not explicitly state them here.

\subsection{Cost-Based Algorithm (CBA)}  \label{sec:CBA}
Since CBA enforces average power constraints per beam, the following additional
constraint to the NUM problem (\ref{eq:SecUtilBegin} to \ref{eq:SecUtilEnd}) is
introduced.

\begin{equation}
\forall b: \bar{P}_{jb}^{m}  \geq \sum_{l}\pi\left(l\right)\sum_k \phi_{jk}^{lm}
c_{jb}\left(k, \bar{P}_j^m\right) \label{eq:CB-SecUtilEnd}
\end{equation}

It includes cost term $c_{jb}\left(k, \bar{P}_j^m\right)$ which represents the
power cost or power consumption of beam $b$ (on PRB $j$) given
scheduling decision $k$ and $\bar{P}_j^m$, which is the total power budget of
PRB $j$. We assume that each beam that is activated gets an equal share of
the available total PRB power $\bar{P}_j^m$. Thus, if $n_{j}(k)$ is the
number of beams that are activated on PRB $j$ if decision $k$ is chosen, the
`cost' of activating beam $b$ on PRB $j$ becomes

\[
c_{jb}(k,\bar{P}_j^m) = \frac{1}{n_j(k)}\bar{P}_j^m,
\] with $\bar{P}_j^m =
\sum_b \bar{P}_{jb}^m$.

To solve the modified problem
(\ref{eq:SecUtilBegin} to \ref{eq:SecUtilEnd} and \ref{eq:CB-SecUtilEnd}), we also
have to modify the gradient scheduler (\ref{eq:GradSched} to \ref{eq:GradSchedVirtRates}).
% \textcolor{red}{\bf\{AU Query: For clarity, the previous sentence was modified. Please confirm if the
% modification is correct.\}} 
%ANSWER: Since the referenced equations actually constitute a single optimization problem, 
%it might be better to use the singular of problem. 
The modified gradient scheduler now chooses the scheduling decision $k^*$
according to

\begin{align} \label{eq:GradSchedWithCost}
k^{\ast}\in\arg\max_{k}[& \nabla_{X}U^m\left(  \bm{\bar{X}}^m\left(t\right)
\right) \sum_b  \bm{\mu}_{jb}^{l(t)}\left(  k \right) \\
&- \sum_b \lambda_{jb}\left(t\right)c_{jb}\left(k,\bar{P}^m_j\right)]\nonumber.
\end{align}

Dual parameters $\lambda_{jb}\left( t\right)$, measuring the deviation of powers
over time from the target power values on a particular beam, are updated
according to the following rule:

 \begin{equation}
  \label{eq:GradSchedWCost-LambdaUpdate}
  \lambda_{jb}\left( t+1\right)=\left[\lambda_{jb}\left( t\right) + \beta_3
 \left(c_{jb}\left(k^\ast,\bar{P}_j^m\right) - \bar{P}_{jb}^m\left(
t\right)\right)\right]^{+}.
 \end{equation}

Average user rates $\bm{\bar{X}}^m\left(t\right)$ are maintained and updated as
in (\ref{eq:GradSchedVirtRates}). The above algorithm can be seen as an application
of the greedy primal dual (GPD) algorithm presented in \cite{Stolyar2005}.

The virtual control plane differs from VSA in the following.
In contrast to the VSA, where every beam is switched on all the time, the
situation is different here. To enable the
calculation of derivatives of the rates with respect to beam powers (needed in
(\ref{eq:CB-VirtSched-SensUpdate})), we introduce scaling factors
$\alpha_{jb}^{m}\left(k\right)$, which scale target beam powers
$\bar{P}_{jb}^{m}$ to powers `costs'
\footnote{To avoid confusion, we denote variables belonging to the real scheduler with lower case symbols,
and variables from the virtual scheduler with corresponding upper case symbols.}
$C_{jb}^{m}\left(k\right)$ that are instantaneously used by the virtual
scheduler. Thus, $\alpha_{jb}^{m}\left(k\right)\bar{P}_{jb}^{m}=C_{jb}^{m}\left(k\right)$.

Given average gains (\ref{eq:avgGains}) as in VSA, the sector controllers
calculate (virtual) user rates given by

\begin{equation}
\label{eq:CB-VirtRates}
R_{ijb}^{m}\left(  k\right)  =\rho\left( F_{ijb}^m(k)\right),
\end{equation}

with

\begin{equation}
F_{ijb}^m(k) = \frac{\alpha_{jb}^{m}\left(k\right)G_{ijb}^{m}\bar{P}_{jb}^{m}
}{\sigma^{2}+ I_{\text{intra}} + I_{\text{inter}}},
\label{eq:F_CBA}
\end{equation}

$I_{\text{intra}} = \sum_{b^{\prime}\neq b}
\alpha_{jb^\prime}^{m}\left(k\right)G_{ijb^\prime}^{m}\bar{P}_{jb^\prime}^{m}$,

and
$I_{\text{inter}} = \sum_{\hat{m}\neq m}\sum_{\hat{b}}G_{ij\hat{b}}^{\hat{m}}
\bar{P}_{j\hat{b}}^{\hat{m}}$.

Virtual average user rates are calculated by CBA as follows:

\begin{equation}
X_i^m = \sum_j\sum_{k}\tilde{\phi}^m_{jk}\sum_b{R^m_{ijb}(k)}.
\end{equation}

Again, $\tilde{\phi}^m_{jk}$ are optimal time fractions of resource usage for
sector $m$; however, in contrast to VSA where those time fractions were calculated
explicitly, CBA uses the approach of OA to determine the virtual average rates (and
implicitly the time fractions) through \textit{virtual scheduling}.
As before, to distinguish real and virtual scheduler, we use capital letters for
virtual scheduler quantities whenever possible. In each TTI, the virtual scheduler performs $n_v$ scheduling
runs. In each run, the following steps are carried out on each PRB $j$:

\begin{list}{\labelitemi}{\leftmargin=1em}
\item We determine the \textit{virtual} scheduling decision $k^\ast$ similar to (\ref{eq:GradSchedWithCost}).
\item We update \textit{virtual} average rates similar to (\ref{eq:GradSchedVirtRates}).
\item We update the \textit{virtual} average power costs for each beam $b$ similar to (\ref{eq:GradSchedWCost-LambdaUpdate}).
\item We update sensitivities for each beam $b$ and sector $\hat{m}$ ($\beta_2>0$ small) according to
\begin{small}
\begin{equation}
D_{jb}^{\left(  \hat{m},m\right)  }\left( \frac{t+1}{n_{v}}\right) =\left(
1-\beta_{2}\right)  D_{jb}^{\left(  \hat{m},m\right)  }\left( \frac{t}{n_{v}}
\right)
 +\beta_{2}\sum_{i=1}^{I_m}\frac{\partial U^{m}\left(  \bm{X}^{m}\right)
}{\partial X_{i}^{m}}\sum_{b^\prime} \frac{\partial R_{ijb^\prime}^{m}\left(
k^{\ast}, \frac{t}{n_v}\right)
}{\partial\bar{P}_{jb}^{\hat{m}}\left(  t\right)  }.
\label{eq:CB-VirtSched-SensUpdate}
\end{equation}
\end{small}
\end{list}
Using (\ref{eq:CB-VirtRates}) to (\ref{eq:F_CBA}), the derivatives in (\ref{eq:CB-VirtSched-SensUpdate}) are given by
\[
\frac{\partial R_{ijb^{\prime}}^{m}}{\partial P_{jb}^{\hat{m}}}=\left\{
\begin{array}
[c]{cc}
\rho^{\prime}\left(  F_{ijb^{\prime}}^{m}\right)  \frac{F_{ijb^{\prime}}^{m}
}{P_{jb^{\prime}}^{\hat{m}}} & \hat{m}=m,b^{\prime}=b\\
-\rho^{\prime}\left(  F_{ijb^{\prime}}^{m}\right)  \frac{\left(
F_{ijb^{\prime}
}^{m}\right)  ^{2}}{P_{jb^{\prime}}^{m}}\frac{G_{ijb}^{m}\alpha_{jb}^m}{G_{ijb^{\prime}
}^{m}\alpha_{jb^{\prime}}^m} & \hat{m}=m,b^{\prime}\neq b\\
-\rho^{\prime}\left(  F_{ijb^{\prime}}^{m}\right)  \frac{\left(
F_{ijb^{\prime}
}^{m}\right)
^{2}}{P_{jb^{\prime}}^{m}}\frac{G_{ijb}^{\hat{m}}\alpha_{jb}^{\hat{m}}}{G_{ijb^{\prime}
}^{m}\alpha_{jb^{\prime}}^{m}} & \hat{m}\neq m
\end{array}.
\right.
\]

The power adaption is then carried out similar to the other algorithms. For each
resource, each sector sums up values $D_{jb}^{\left(  \hat{m},m\right)}$
 from all sectors and increases the power level on the beam with largest
positive sum while decreasing the power level on the beam with largest negative
sum. However, CBA adapts only power constraints per beam, not actually used beam
powers.

\begin{figure}[h!]
\centering
\includegraphics[width=0.8\textwidth]{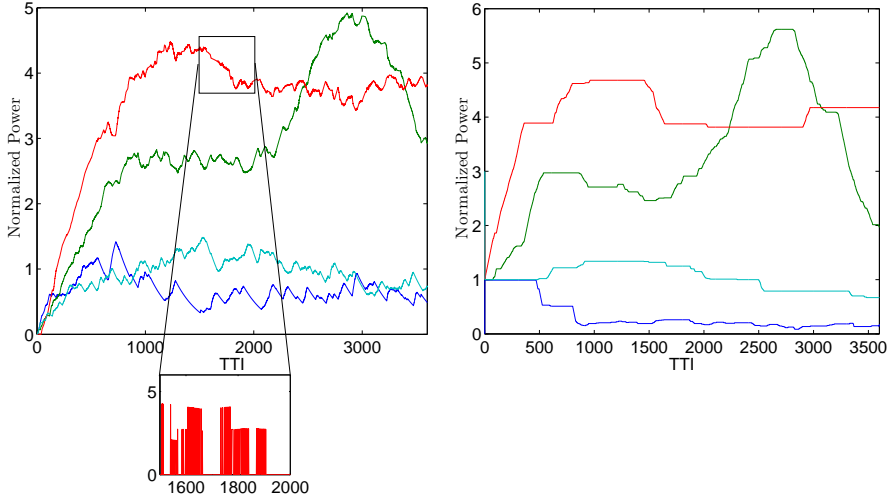}
\caption{(Averaged) power trajectories of `real' scheduler (left) and target powers given by virtual model (right).
The power allocations of four exemplary beams are compared. The `zoom' indicates the averaging of powers over time.
The four curves (red, green, light blue, and dark blue) indicate the averaged per-beam powers over time of the beams (here, a codebook with four entries was used)
of an arbitrary chosen PRB, both for the 'real' scheduler (left), and for the virtual layer (right)).}
\label{fig3}
\end{figure}

Figure~\ref{fig3} gives a sketch of power trajectories using the
simulation environment described below, in Section~\ref{Sec:Sim}.
To find out whether CBA really holds the average power constraints,
we exponentially average instantaneously used powers per beam with the same
time constant used for scheduling and compare the result to the target power values
determined by the virtual model.
The left side of Figure~\ref{fig3} shows the averaged powers, as
actually used by the `real' scheduler, while
the right side shows the target power values determined by the virtual
scheduling procedure.
Note that since scheduling is opportunistic, instantaneously, the power levels
fluctuate highly and beam powers can differ from the target values (or a beam
can be completely turned off). This is illustrated by the `zoomed-in image'
in Figure~\ref{fig3} (left), where actual powers without averaging
are shown. It turns out that \textit{on average}, the power constraints are kept
remarkably well.

Apart from the intuitive benefits of instantaneously allowing opportunistic
scheduling, we observe
that from a practical point of view, average power constraints are further
justified since hybrid automatic repeat request (HARQ) coding is essentially
performed over multiple successive transmissions.

In all three presented algorithms, naturally, questions arise regarding
differentiability.
Note that the problem to be solved by each sector controller is given in
(\ref{eq:VSAoptimization}).
Due to the maximum operator, it might not be differentiable everywhere even with
the utility function being differentiable.
Therefore, let us define a slightly modified version
of problem (\ref{eq:VSAoptimization}), given by

\begin{equation}
\label{eq:3}f_{\varepsilon}\left(  R\right)  :=\max_{\phi_{ijb}^{m}}\sum_{i}
U^m\left(  \sum_{j}\sum_{b}\left(  \phi_{ijb}^{m}\right)  ^{1-\varepsilon
}R_{ijb}^{m}\right)  .
\end{equation}

One can show the following:

\begin{mytheorem}\label{Thm:2}
Let $0<\varepsilon<1$ be finite and $U^m$ be an increasing concave
utility function, defined in $(0,\infty)$. Then, the family of functions
$f_{\varepsilon}\left(R\right)$ (defined by (\ref{eq:3})) with $R_{ijb}\geq c>0$  ($\forall i,j,b$)
is differentiable everywhere and converges for
any sequence $\varepsilon_{n}\rightarrow0$ to $f$ in (\ref{eq:VSAoptimization})
(which is continuous) in the uniform sense.
\end{mytheorem}

\begin{myproof}
The proof can be found in Appendix 1.
\end{myproof}

By Theorem~\ref{Thm:2}, we can replace our utility function with a smooth,
uniformly convergent approximation, which can be locally maximized in the
power control loop. Note that this replacement is already incorporated
in the calculation of sensitivities (Equation \ref{eq:CalcSensitivities}).

\section{An alternating optimization based approach} \label{sec:AO}

Given the distributed nature of the above presented algorithms, the question
arises: Can a centralized controller do better?
Therefore, in the following,
we want to compare the algorithms of Section~\ref{sec:distributedAlgs} with a
centralized solution.

The optimization problem which the virtual controller has to solve is given by

\begin{align}
\max_{\bm{P},\bm{\Phi}}\quad
&\sum_{m=1}^M\sum_{i=1}^{I_m}\log\left(\sum_{j=1}^J\sum_{b}\phi_{ijb}^m
R_{ijb}^m(\bm{P})\right) \label{eq:optproblem_central} \\
\text{s.t.}\quad& \forall b,j,m:\quad \sum_{i} \phi_{ijb}^m \leq 1
\label{eq:centralprobphiconstr} \\
& \forall m: \quad \sum_b \sum_{j} P_{jb}^m \leq P_{\text{max}},
\label{eq:centralprobpowconstr}
\end{align}

with $R_{ijb}^m(\bm{P})=\log\left(1+F_{ijb}^m(\bm{P})\right)$ and $F_{ijb}^m$
defined in (\ref{eq:F}).
As described in Section~\ref{sec:distributedAlgs}, the power allocation problem
is so far solved using a distributed gradient ascent procedure.
However, when we allow a centralized controller for the network,
we can instead solve
(\ref{eq:optproblem_central} to \ref{eq:centralprobpowconstr}) directly each time a
power update is desired and use the resulting power allocation
directly for actual resource allocation.

Obviously, problem (\ref{eq:optproblem_central} to \ref{eq:centralprobpowconstr})
is highly non-convex. In the following, we try to solve the problem by alternating the optimization in
$\bm{P}$ (holding $\bm{\Phi}$ constant) and $\bm{\Phi}$ (holding $\bm{P}$
constant). We therefore have a scheduling sub-problem and a power allocation (PA)
sub-problem. The overall procedure is summarized in Algorithm~\ref{alg:AO}.

\begin{algorithm}[H]
   \caption{\bf Alternating optimization-based virtual layer}
   \begin{algorithmic}[1]
      \STATE initialize counter $\tau=0$, select feasible $\bm{\Phi}(0)$ and
$\bm{P}(0)$ arbitrarily
      \REPEAT
      \STATE $\tau\leftarrow\tau +1$
      \STATE solve scheduling sub-problem using fixed $\bm{P}(\tau-1)$ and obtain
$\bm{\Phi}(\tau)$
      \STATE solve SCA-based power allocation sub-problem using fixed
$\bm{\Phi}(\tau)$ and obtain $\bm{P}(\tau)$ (Algorithm~\ref{alg:SCA_PC})
      \UNTIL converged
      \STATE use $\bm{P}(\tau)$ to update power allocation in network
    \end{algorithmic}
   \label{alg:AO}
\end{algorithm}

Since the scheduling sub-problem is convex (cf. Lemma~\ref{lemma:optproblemphi}),
we only have to care about the power allocation sub-problem.
We try to tackle this problem by a \textit{successive convex approximation}
(SCA) approach similar to \cite{Papandriopoulos2009}.
The sub-problem in $\bm{P}$ is still highly non-convex. However, using Lemma
\ref{lemma:pow_alloc_subprob_convexify}, we obtain a convexified version of the
power allocation sub-problem.

\begin{mylemma} \label{lemma:optproblemphi}
 With constant $\bm{P}$, optimization problem
(\ref{eq:optproblem_central}-\ref{eq:centralprobpowconstr}) is a convex
optimization problem in $\bm{\Phi}$.
\end{mylemma}

\begin{myproof}
 The proof follows since non-negative weighted addition and scalar composition
preserve concavity \cite{boyd2004}.
\end{myproof}

\begin{mylemma} \label{lemma:pow_alloc_subprob_convexify}
Using a concave lower bound (assuming appropriately chosen constants) to the
user rates, given by
\begin{equation}
\tilde{R}_{ijb}^m:= \alpha_{ijb}^m \log(F_{ijb}^m) + \beta_{ijb}^m \leq
{R}_{ijb}^m, \label{eq:ConcaveLowerBound}
\end{equation}
(with equality holding when approximation constants are chosen as
$\alpha_{ijb}^m=\frac{F_{ijb}^m}{1+F_{ijb}^m} $ and $\beta_{ijb}^m =
\log(1+F_{ijb}^m)-\alpha_{ijb}^m\log(F_{ijb}^m)$) and a logarithmic change of
variables given by $\tilde{P}_{jb}^m = \log({P_{jb}^m})$,
the modified PA sub-problem

\begin{align}
\max_{\bm{\tilde{P}}}\quad
&\sum_{m=1}^M\sum_{i=1}^{I_m}\log\left(\sum_{j=1}^J\sum_{b}\phi_{ijb}^m
\tilde{R}_{ijb}^m (\tilde{\bm{P}})\right)  \label{eq:PAsubproblemobjective}\\
& \forall m: \quad \sum_b \sum_{j} e^{\tilde{P}_{jb}^m} \leq P_{\text{max}}
\label{eq:PAsubproblempowconstr}
\end{align}

is a convex optimization problem.
\end{mylemma}

\begin{myproof}
The proof can be found in Appendix 2.
\end{myproof}

The SCA procedure is summarized in Algorithm~\ref{alg:SCA_PC}.

\begin{algorithm}[H]
   \caption{\bf SCA-based power allocation sub-problem}
   \begin{algorithmic}[1]
      \STATE initialize counter $\tau=0$, $\bm{\alpha}(0)=\bm{1};\quad
\bm{\beta}(0)=\bm{0}$
    \REPEAT
    \STATE solve
(\ref{eq:PAsubproblemobjective} to \ref{eq:PAsubproblempowconstr}), obtain
$\tilde{\bm{P}}^{\ast}(\tau)$
    \STATE tighten approximation ($\forall i,j,b,m$):
\begin{small}
    \begin{align}
     \alpha_{ijb}^m(\tau+1)=&\frac{F_{ijb}^m(\bm{P}^{\ast}(\tau))}{1+F_{ijb}^m(\bm{P}
^{\ast}(\tau))}  \label{eq:SCA_tighten1}\\
     \beta_{ijb}^m(\tau+1)=&\log\left(1+F_{ijb}^m\left(\bm{P}^{\ast}
\left(\tau\right)\right)\right) \nonumber \\
     &-\alpha_{ijb}^m\log\left(F_{ijb}^m\left(\bm{P}^{\ast}
\left(\tau\right)\right)\right) \nonumber
    \end{align}
\end{small}
    \STATE $\tau \Leftarrow \tau+1$
      \UNTIL converged
    \end{algorithmic}
   \label{alg:SCA_PC}
\end{algorithm}

The algorithm is initialized as described in the first step of Algorithm
\ref{alg:SCA_PC}, which is equivalent to a high signal-to-interference-plus-noise ratio (SINR) approximation
(which can be seen when applying this initialization in (\ref{eq:ConcaveLowerBound})). The high-SINR
approximation assumes that for large SINR, $\log(1+\mathrm{SINR})\approx \log(\mathrm{SINR})$.
This is a common initialization for this kind of algorithm \cite{6129545, Papandriopoulos2009}.
Each iteration of the algorithm comprises two steps, a maximize step and a tighten-step.
In the maximize step, we find a solution to the current convexified version
(\ref{eq:PAsubproblemobjective} to \ref{eq:PAsubproblempowconstr}) of the power
control problem. This solution is then used in the tighten step to update
the convex approximation parameters $\bm{\alpha}$ and $\bm{\beta}$ for each link
according to (\ref{eq:SCA_tighten1}). The  algorithm converges when the tighten step (\ref{eq:SCA_tighten1})
does not produce any (significant) changes. Being based on an inner approximation framework by Marks and Wright
\cite{marks1978}, it can be shown that Algorithm~\ref{alg:SCA_PC} converges at least to a
KKT point of the PA sub-problem.

\section{Approaching Global Optimality: Comparison With Branch-And-Bound} \label{sec:BNB}

It is known that the underlying optimization problem is non-convex; thus, the
gradient ascent-based algorithms presented in Section~\ref{sec:distributedAlgs}
as well as the alternating optimization-based algorithm of Section~\ref{sec:AO}
will most likely converge to a local maximum. Thus, although simulation results
(cf. Section~\ref{sec:simulations}) show already high gains in utility, the
question remains how good the solution found actually is, that is how much of
the achievable performance gains is actually realized?

To simplify the analysis in this section, we restrict ourselves to the single-antenna case
(which reduces all algorithms to MGR in \cite{Infocom09}). The main difficulty of the analysis
is that even in this simple setting, the underlying problem
\begin{align}
f(\bm{P},\bm{\Phi}) = \max_{\bm{P},\bm{\Phi}}\quad
&\sum_{m=1}^M\sum_{i=1}^{I_m}\log\left(\sum_{j=1}^J\phi_{ij}^m R_{ij}^m\right)
\label{eq:orgproblem} \\
\text{s.t.}\quad& \forall j,m:\quad \sum_{i} \phi_{ij}^m \leq
1\label{eq:orgprobphiconstr} \\
& \forall m: \quad \sum_{j} P_{j}^m \leq P_{\text{max}}
\label{eq:orgprobpowconstr}
\end{align}

with $R_{ij}^m=\log\left(1+\frac{G_{ij}^{m}P_j^{m}}{\sigma^2+\sum_{m^\prime \neq
m} G_{ij}^{m^\prime}P_j^{m^\prime}}\right)$, is highly non-convex (even in the
two-sector, two-user case).
Moreover, since this is essentially a joint optimization in
time fractions $\bm{\Phi}$ and power values $\bm{P}$, the complexity is
significantly higher than with power control only.
To gain insight into the deviation from the optimal solution of
(\ref{eq:orgproblem} to \ref{eq:orgprobpowconstr}), we compare our algorithm's
performance in the simplified setting with a (near) optimal solution based on
BNB. Although computationally very expensive, it gives us a good impression on
how much of the achievable performance is actually reached.
We assume, without loss of generality, that the maximum sum power available in
each cell in (\ref{eq:orgprobpowconstr}) is normalized to one.

Branch-and-bound is a standard algorithm for global optimization, which creates
a search tree where at each node, an upper and a lower bound to the problem are
evaluated. Details can be found for example in \cite{GlobOpt}.
It is based on constantly sub-dividing the feasible parameter region and for
each node of the resulting tree calculating
the upper and lower bounds to the objective function.
For ease of notation, we will combine all parameters to our objective function in one vector
$\bm{\hat{P}} = \left( p_1,\ldots,p_n \right)$.
Let $\mathcal{P}^{(0)}$ denote our initial parameter region.
This region forms a convex $n$-dimensional polytope with $V$ extreme points (or vertices)
$\bm{\hat{P}}_v$. We collect the corner points of polytope $\mathcal{P}^{(0)}$ in set
$\mathcal{V}^{(0)}$.
Consequently, the parameter region associated with node $l$ is denoted as
$\mathcal{P}^{(l)}$ with associated vertices $\mathcal{V}^{(l)}$.

\begin{figure}[h!]
\centering
\includegraphics[width=0.6\linewidth]{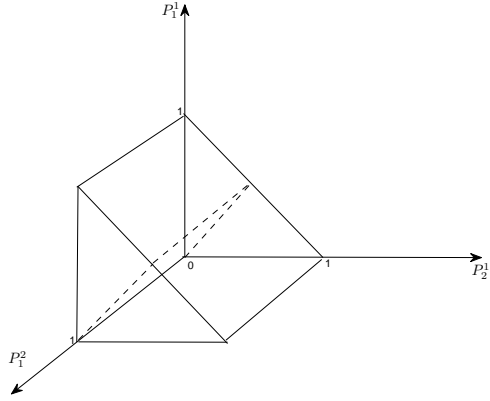}
\caption{3-dimensional polytope with indicated branching}%
\label{fig4}%
\end{figure}

Due to the structure of the constraints, we have $\frac{n}{2}$ parameter pairs
with independent constraints. Each of the parameter/constraint pairs form a
(two-dimensional) simplex.
When branching, we split the parameter region of node $l$, $\mathcal{P}^{(l)}$,
in sub-regions $\mathcal{P}^{(l+1)}$ and $\mathcal{P}^{(l+2)}$ by splitting the
longest edge among the edges of all ($\frac{n}{2}$) simplexes together.
Doing so, we divide the parameter region into two sub-regions of equal size.
Figure~\ref{fig4} visualizes the above-stated procedure by showing the
simplex created by two interdependent power values together with an independent
third dimension. The dashed line indicates the splitting into two sub-polytopes.

Below, we explain how the upper and lower bounds to the sub-problems are found.
It is well known that the Shannon rate can be written as a difference of concave functions
$f_{ij}^{m}(\bm{P})-g_{ij}^{m}(\bm{P})$, with $f_{ij}^{m}(\bm{P}) = \log\left(\sigma^2 +\sum_{m^\prime=1}^M
G_{ij}^{m^\prime}P_j^{m^\prime}\right)$ and $g_{ij}^{m}(\bm{P})=\log\left(\sigma^2+\sum_{m^\prime \neq m}
G_{ij}^{m^\prime}P_j^{m^\prime}\right)$.
Using this, our non-concave objective function (\ref{eq:orgproblem}) can be rewritten as
\begin{equation}
\hat{F}(\bm{\hat{P}})=\sum_{m=1}^M\sum_{i=1}^{I_m}\log\left(\sum_{j=1}^J
\phi_{ij}^{m}f_{ij}^{m}(\bm{\hat{P}}) - \phi_{ij}^{m}g_{ij}^{m}(\bm{\hat{P}})\right)\label{eq:uboundorgfunc}.
\end{equation}

Unfortunately, $\phi_{ij}^{m}f_{ij}^{m}(\bm{\hat{P}})$ (and thus also
$\phi_{ij}^{m}g_{ij}^{m}(\bm{\hat{P}})$) is neither convex nor concave (note
that $f(a,b)=a\cdot\log(1+b)$ is not concave).
However, we can define the concave function(s):\\
$
 \tilde{f}_{ij}^{m}(\bm{\hat{P}}) =
\phi_{ij}^{m}\log\left(\sigma^2+\frac{1}{\phi_{ij}^{m}}\sum_{m^\prime=1}^MG_{ij}
^{m^\prime}P_j^{m^\prime}\right)
$,
 and in the same way $\tilde{g}_{ij}^{m}(\bm{\hat{P}})$.
Using this, (\ref{eq:uboundorgfunc}) can be upper bounded by
\begin{equation}
\sum_{m=1}^M\sum_{i=1}^{I_m}\log\left(\sum_{j=1}^J
\tilde{f}_{ij}^{m}(\bm{\hat{P}}) - \tilde{g}_{ij}^{m}(\bm{\hat{P}})\right).
\label{eq:uboundphitrick}
\end{equation}

Equation \ref{eq:uboundphitrick} is still not concave since it includes a sum of a
concave function ($\tilde{f}_{ij}^{m}(\bm{\hat{P}})$) and a convex function ($-\tilde{g}_{ij}^{m}(\bm{\hat{P}})$).
To get a concave objective function, we replace $\tilde{g}_{ij}^{m}(\bm{\hat{P}})$ by it's \textit{convex envelope}
(cf. \cite{GlobOpt}), defined as follows:

Let $\bm{P}_1,\ldots,\bm{P}_V$ be the vertices of a polytope $\mathcal{P}$. The convex envelope
$\gamma(\bm{P})$ of a concave function $g(\bm{P})$ over $\mathcal{P}$ can be expressed as
\[
\gamma(\bm{P}) = \min_{\bm{\lambda}}\sum_{v=1}^V\lambda_v g(\bm{P}_v),
  \]
subject to $\sum_{v=1}^V \lambda_v \bm{P}_v = \bm{P}$,
$\sum_{v=1}^V\lambda_v=1$, and $\lambda_v\geq 0$.
Using this definition, the (convex) optimization problem we have to solve in
order to obtain our upper bound (at node $l$) is given by
\vspace*{-3pt}\begin{align*}
F^U=\max_{\bm{\lambda}}&\sum_{m=1}^M\sum_{i=1}^{I_m}\log\left(\sum_{j=1}^J
\tilde{f}_{ij}^{m}(\bm{\hat{P}}) - \sum_{v=1}^V\lambda_v
g(\bm{\hat{P}}_v)\right) \\
  \text{s.t.}\quad& \bm{\hat{P}} = \sum_{v=1}^V \lambda_v \bm{\hat{P}}_v,\;
   \sum_{v=1}^V\lambda_v=1,\;
   \lambda_v\geq 0\; (\forall v),
\end{align*}
\vspace*{-3pt} with $\bm{\lambda} = (\lambda_1,\ldots,\lambda_V)$,
$\bm{\hat{P}}\in\mathcal{P}^{(l)}$, $\bm{\hat{P}}_v\in\mathcal{V}^{(l)}$, and
$V$ being the number of vertices collected in $\mathcal{V}^{(l)}$.\\
We use three different approaches to obtain a lower bound at node $l$.
The first lower bound is obtained by taking the maximum of the objective function values at each of
the corner points of the respective parameter region. Second, we evaluate the optimal parameter
vector from calculating the upper bound. Third, we use a standard solver to compute a (local)
optimum of the original non-convex problem. If one of the three methods leads to a higher value,
the current global lower bound is replaced.

\section{Numerical Evaluation} \label{sec:simulations}

In this section, we present numerical results to evaluate the distributed
algorithms of Section~\ref{sec:distributedAlgs} as well as the
centralized and BNB-based solutions of Section~\ref{sec:AO} and Section
\ref{sec:BNB}, respectively.

\subsection{Distributed Algorithms (System-Level Simulations)} \label{Sec:Sim}

In order to compare the performance of the three algorithms in a setting close
to practice, we conduct system-level simulations based on LTE.

\subsubsection{Simulation Setup}

The general simulation setup can be found in Table~\ref{tab2}.
We employ a grid of seven hexagonal cells, each comprising three sectors (to ensure equal interference conditions
for each sector, a wrap-around model is used at the system borders). Users are distributed randomly over
the whole area; thus although we have, say, 210 user in total, leading to an average of ten users per
sector, some sectors have a higher load than others.
A detailed description of the precoding codebook can be found in \cite{BeamDis2007}.
As channel model, we use the WINNER model \cite{WINNER5.4}, Scenario C2.
All simulations are performed with realistic link adaptation based on mutual information effective SINR mapping
(MIESM); moreover, we use explicit modeling of HARQ (hybrid automatic repeat request) using chase combining.
For efficiency reasons, we simulate a limited number of PRBs ($J=8$) compared to a real-world system;
however, the results are expected to scale to a higher number of PRBs.

%== Inserted Table 2
%== Table 2 ==
%\vspace*{-18pt}
\begin{table}[!h]
\centering
\caption{\bf General simulation setup}
%\textcolor{red}{\bf\{AU Query: Please confirm if the modifications done are correct.\}}}
%ANSWER: The modifications are correct. 
\label{tab2}
\begin{tabular}[c]{lc}
\hline
\textbf{Parameter}                      & \textbf{Value}\\
\hline
Number of sectors ($M$)                 & $21$\\
Total number of terminals ($I$)         & $105$ to $315$\\
Mobile terminal velocity                & $0$ km/h, $3$ km/h\\
Number of PRBs ($J$)                    & $8$\\
Number of beams ($B$) (coordination)    & $4$\\
Number of beams ($B$) (baseline)        & $8$\\
Base station antennas $(n_\mathrm{T})$  & $4$\\
Number of terminal antennas ($n_\mathrm{R}$) & $1$\\
Simulation duration                     & $10,000$ TTI\\
Power adaption step size ($\Delta$)     & $0.5\%$ of the initial power\\
Traffic model                           & Full buffer\\
\hline
\end{tabular}\\
\footnotesize{This table summarizes the main parameters that we use in the system-level simulations.}
\end{table}

\subsubsection{Baseline Algorithm} \label{sec:baseline}
As baseline, we use a non-coordinative scheduling algorithm called greedy
beam distance (GBD) algorithm, with a codebook size of 3 bits (eight beams).
GBD requires feedback from each user, comprising a CDI and a channel quality information (CQI) value.
Note that the CQI is only an estimate of the users' SINR values in case the user alone is scheduled on the PRB
(with full power), since the scheduling decisions cannot be known in advance. Thus, it does not contain
intra-sector interference. Moreover, it contains only an estimate of the inter-sector interference.
On each PRB, the users are greedily scheduled on their best beams (using their
proportionally fair weighted CQI feedback as utility). Thereby, a minimum
beam distance has to be kept, in order to minimize the interference between
users scheduled on the same PRB. This distance is based on a geometrical
interpretation of beamforming. It means that given a user is scheduled on a certain beam, adjacent beams (up to a
certain `distance') are blocked and users that reported one of those beams are
excluded from the list of scheduling candidates for the respective PRB. We use
a minimum distance of 3, that is, with the used eight-beam codebook at most three
users can be scheduled on the same PRB. Of course, no adaptive power allocation is performed;
the power is distributed equally among the PRBs (and further among the thereon scheduled users).

\subsubsection{Simulation Results}
We use two essential performance metrics for our comparison. First, since we
have a system-wide proportional fair utility, we compare
the geometric mean of average user rates (GAT) as a measure of the increase of
the \textit{overall} utility. The equivalence of maximizing GAT and sum utility can be seen when observing
that the geometric mean of some set of values is equal to the exponential average of the logarithm of the values.
Therefore, the sum of logarithms is maximized precisely if the geometric mean
is maximized. Second, the performance of cell-edge users, which we measure by the 5\% quantile of
average user throughputs, is a natural benchmark for each distributed base station
coordination algorithm.

\begin{figure}[h!]
\centering
\includegraphics[width=0.7\textwidth]{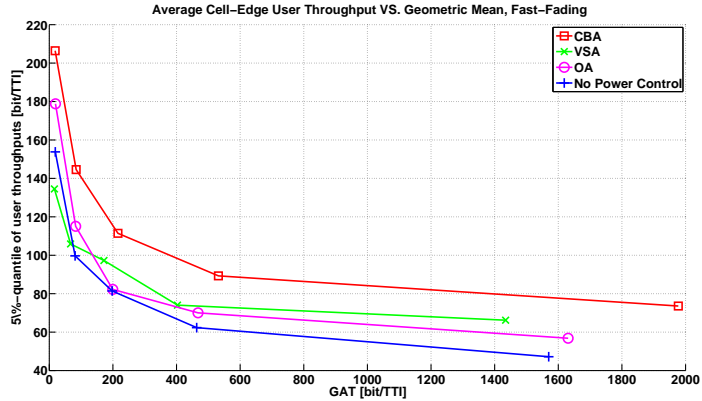}
\caption{Cell-edge user throughput vs{.} GAT in a fading environment.
%\textcolor{red}{\bf\{AU Query: Please confirm if the modified titles of Figures 5 and 6 are correct.\}}
%ANSWER: Yes, the modified titles are correct. 
The averaged cell-edge users' throughputs vs. the geometric mean of users'
throughputs are plotted for a fast-fading scenario with users moving at 3 km/h. Thereby, the proposed three
distributed algorithms are compared with a baseline algorithm without coordination.
For each algorithm, the total number of users in the network is varied. In particular,
the markers in the figure, from left to right, represent an average number of 5, 8, 10, 12, and 15
users per sector. This corresponds to total number of 105, 168, 210, 252, and 315 users in the network, respectively.}
\label{fig5}
\end{figure}

\begin{figure}[h!]
\centering
\includegraphics[width=0.7\textwidth]{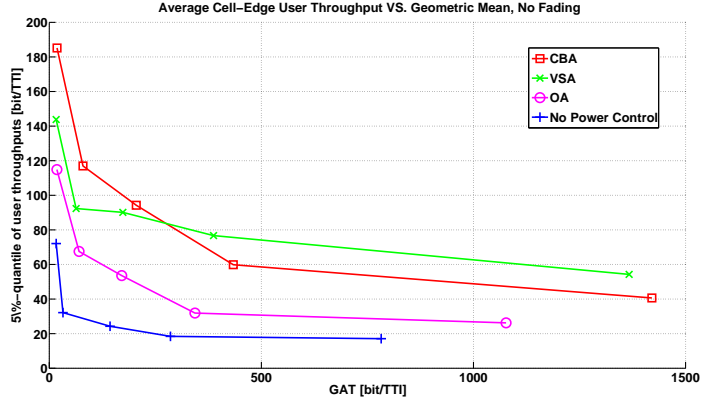}
\caption{Cell-edge user throughput vs{.} GAT in a static environment.
The averaged cell-edge users' throughputs vs. the geometric mean of users'
throughputs are plotted for a scenario without user mobility. Thereby, the proposed three distributed algorithms
are compared with a baseline algorithm without coordination. For each algorithm, the total number of users in the
network is varied. In particular, the markers in the figure, from left to right, represent an average number of
5, 8, 10, 12, and 15 users per sector. This corresponds to total number of 105, 168, 210, 252, and
315 users in the network, respectively.}
\label{fig6}
\end{figure}

Figure~\ref{fig5} shows cell-edge user throughput over GAT for the
evaluated algorithms with user mobility (and therefore
with fast fading), while Figure~\ref{fig6} depicts the simulation
results in a setting without user mobility.
We compare the three approaches of Section~\ref{sec:distributedAlgs} to the
uncoordinated baseline of Section~\ref{sec:baseline}.
Thereby, for each algorithm, the performance is evaluated for different average
number of users per sector. More precisely, from left to right,
the markers in the figure represent an average number of 5, 8, 10, 12, and 15
users per sector (corresponding to total numbers of 105, 168, 210, 252, and 315 users in the network, respectively).
Obviously, increasing values on the ordinate corresponds to increased fairness (with respect to our 5\% quantile metric),
while higher values on the abscissa correspond to a higher sum utility (with respect to our GAT metric).
In both figures, we observe that a higher number of users lead
to increased sum utility (an effect called multiuser diversity) while the
cell-edge users (represented by the 5\% of users with lowest throughput) suffer
from a reduced performance (due to an increased competition for resources).

In case of user mobility (Figure~\ref{fig5}), we observe that CBA clearly outperforms the other algorithms.
In fact, for every simulated user density, either the 5\% fairness (low user densities) or both metrics are improved
(higher user densities). For example, in the case of 210 users in the network (corresponding to the third marker on the
curves in Figure \ref{fig5}), the CBA algorithm improves the performance with respect to GAT by about 10\% and in
cell-edge user throughput by more than 35\%. It can also be observed that the VSA algorithm works best at high
user densities, while in case of only a few users in the cell, the OA algorithm (which requires the least
overhead with respect to additional feedback and signaling) and even the no-power control baseline
show a better performance, at least with respect to the cell-edge performance.
The decreased GAT shown by VSA in the fading case is basically the cost of leaving no freedom to the schedulers for
opportunistic scheduling, but to strictly specify the powers to be used for all beams and time instances.

Figure~\ref{fig6} depicts the simulation results in a setting without user mobility. We can see
that OA offers already quite high performance gains, both in GAT and in cell-edge user throughput,
although being the least complex algorithm. However, the performance is even further improved by the other two
algorithms. VSA can improve especially the gains of cell-edge users, for example, in the case of 210 users in the network
to more than 200\% compared with no-power control. However, CBA again shows the best performance,
significantly improving the global utility compared with VSA (and even more compared with no-power control.)
Again, this gain increases with increasing number of users in the network.

\begin{figure}[h!]
\centering
\includegraphics[width=0.7\linewidth]{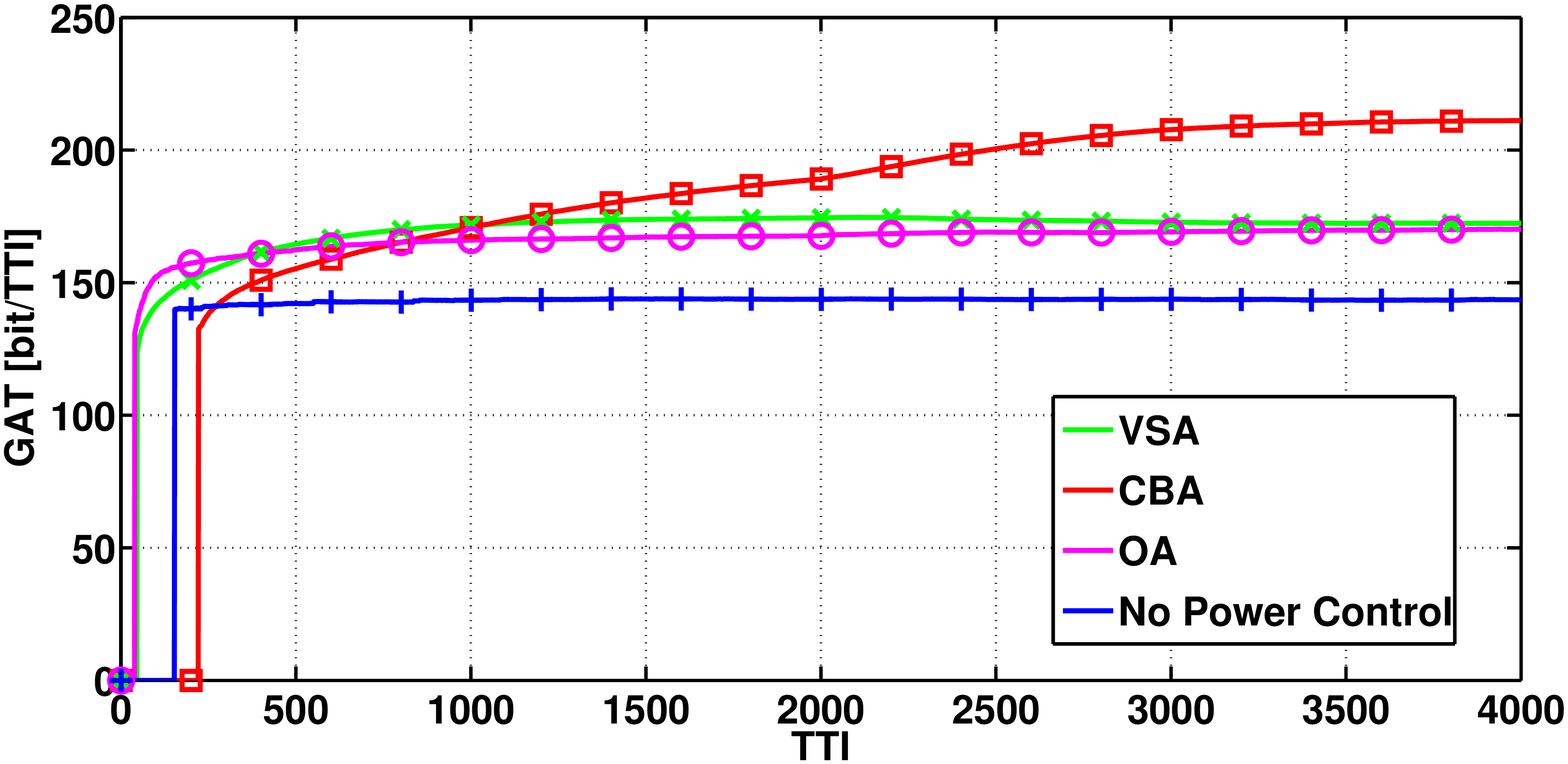}
\caption{GAT over time in a static environment.
The geometric mean of users' throughputs is plotted over the first 4,000 TTIs of the simulation for a scenario
without user mobility. Thereby, the proposed three distributed algorithms are compared with a baseline algorithm
without coordination.}%
\label{fig7}%
\end{figure}

\begin{figure}[!h]
\centering
\includegraphics[width=0.7\linewidth]{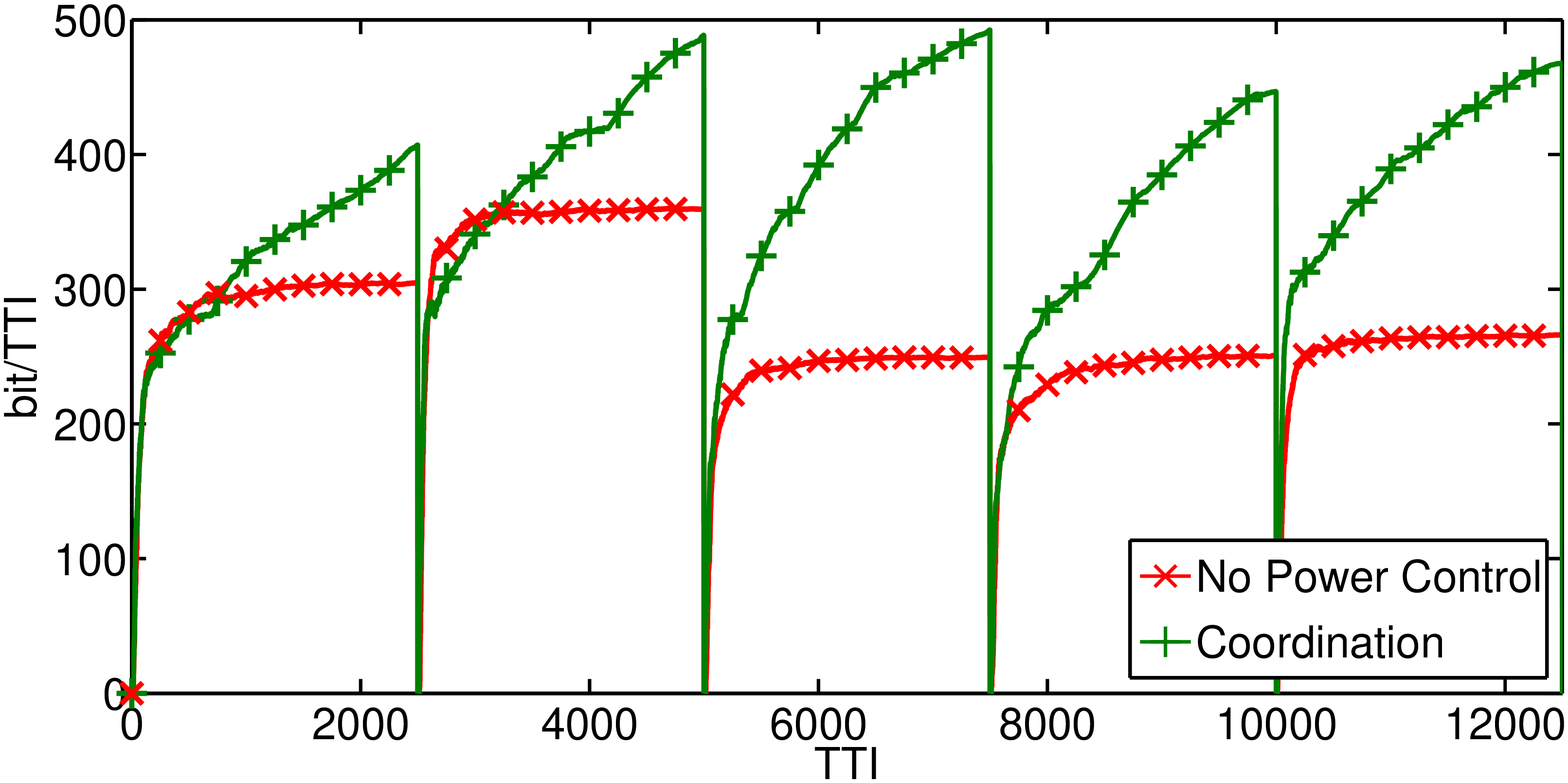}
\caption{Cell-edge user throughput over time for consecutive drops, no user mobility. The green line represents the
CBA algorithm, while the red line indicates the baseline without power control. Multiple consecutive drops with
a duration of 2,500 TTIs are shown.}
\label{fig8}
\vskip-5pt
%\hrulefill
\end{figure}

Besides the performance, the speed of convergence is of large practical interest.
To illustrate this, Figure~\ref{fig7} shows the GAT metric for all algorithms over time for the case of stationary users
(with 210 users in total). It can be seen that the fastest convergence is achieved by the OA and VSA algorithms,
while the CBA algorithm, which in the end achieves the highest GAT, needs about 3,000 TTIs to converge.
This is expectable, since the scheduler has to follow the power values obtained from the virtual layer only on average;
thus, it always `lacks behind' the adaptation of the virtual model. Regardless of which algorithm performs best, it is of
general importance that distributed power control algorithms do not only work in a specific environment but can adapt to
changing network conditions (which, of course, is also related to the notion of convergence).
Figure~\ref{fig8} demonstrates this autonomous adaptation ability by showing the cell-edge user throughput over
time (exemplary for CBA in the static case) for multiple consecutive drops, where, in each drop, a new user location and
assignment pattern are generated. It can be observed that the the performance is improving very fast compared to
the baseline.

\subsection{Performance of centralized solution based on AO}

Subsequently, we investigate the performance of the centralized scheme based on
alternating optimization (AO), as introduced in Section~\ref{sec:AO}.
For this, we compare the performance with that of the distributed solution. More precisely, 
we use the VSA algorithm of Section~\ref{sec:VSA}. Note that the only
difference between the subsequently investigated algorithms is the power
adaptation procedure. While in the distributed solution (in subsequent figures denoted as
\textit{VSA}), the power is adapted gradually based on
the exchange of sensitivities, in the centralized solution (in subsequent
figures denoted as \textit{AO}), the solution to the optimization problem in the
virtual layer is directly used to globally update the power allocation.
We are predominantly interested in the convergence speed of the geometric mean of
average user rates in the different approaches and the relative performances in different
signal-to-noise ratio (SNR) regimes. Moreover, we investigate both a static setting without mobility and
a setting where a user moves at a velocity of 3 km/h.

\begin{figure}[h!]
\begin{center}
\includegraphics[width=0.7\linewidth]{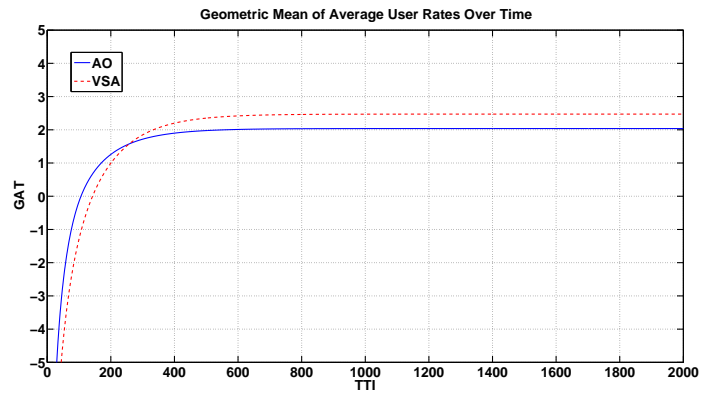}
\end{center}
\caption{Comparison of the centralized and distributed virtual control without mobility.
The geometric mean of the average user rates is shown in a scenario without
mobility for the VSA algorithm (red) and the centralized baseline (blue).}%
\label{fig9}%
\end{figure}

In Figure~\ref{fig9}, we compare the behavior of the AO and the
distributed approach for constant channels.
In general, it can be observed that the convergence speed of the average rates using
AO is slightly faster. Often, both algorithms converge to the same solution;
however, this is not necessarily always the case (as shown in Figure \ref{fig9}).

\begin{figure}[h!]
\begin{center}
\includegraphics[width=0.7\linewidth]{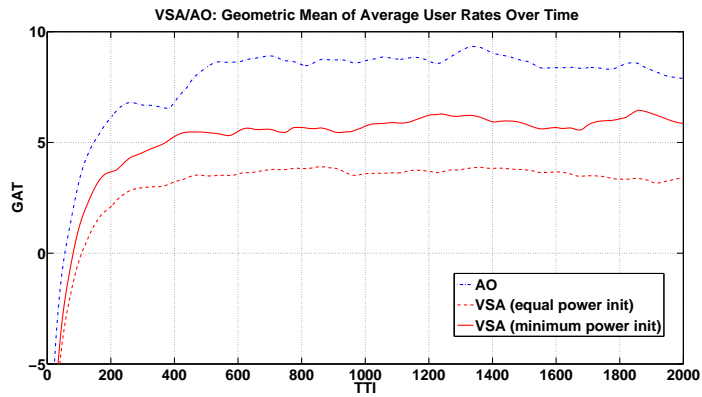}
\end{center}
\caption{Comparison of centralized and distributed virtual control at moderate mobility.
The geometric mean of the average user rates is shown in a fast-fading scenario.
The centralized baseline (blue) is compared to the VSA algorithm (red) with
different initial power allocations.}%
\label{fig10}%
\end{figure}

Figure~\ref{fig10} compares the performance of AO-based and gradient
ascent-based solution in a setting with moderate user mobility.
It can be observed that the convergence speed of the centralized scheme is
still slightly higher than in the distributed case, and in most cases, the
centralized solution outperforms the gradient-based approach
(however, at a much higher complexity).
An interesting effect that can be observed is that the influence of the initial
state on the outcome of the distributed solution is significant. This is also depicted in
Figure~\ref{fig10}. Note that the only difference between the solid and the dashed red line in
Figure~\ref{fig10} lies in the initial power values. In the first case, we start at a minimal power assigned
to all sub-carriers (note that for technical reasons, it is not possible to assign exactly zero power to the resources)
and in the latter case, we start at an allocation where power is distributed evenly among the resources.

\begin{figure}[h!]
\begin{center}
\includegraphics[width=0.7\linewidth]{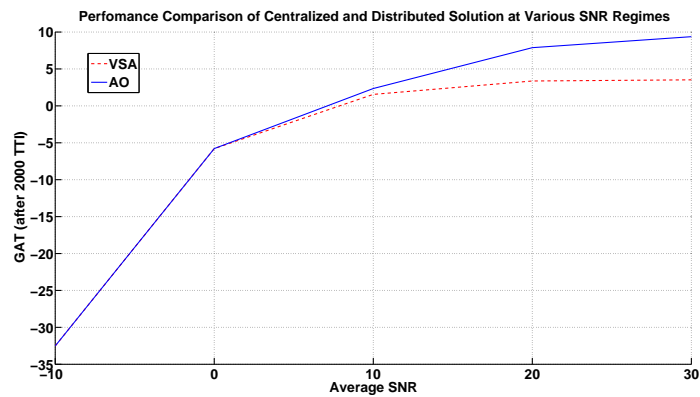}
\end{center}
\caption{Performance at different SNR regimes.
VSA algorithm (red) and centralized algorithm (blue) are compared in terms of
geometric mean of average user throughputs vs. average SNR.}%
\label{fig11}%
\end{figure}

In Figure~\ref{fig11}, it can be observed that the gains from the
centralized solution are higher in high-SNR regimes.
While at low SNRs, when the system is essentially noise-limited, both approaches
converge to the same solution; in high-SNR regimes, when the system is
essentially interference-limited, the direct AO-based optimization outperforms
the gradual power adaptation.

In summary, the centralized solution has some clear advantages. First, it is
invariant to the initialization of the system and is thus able to
adapt very fast to changing environmental conditions. Second, especially when
the channels are varying fast and the SNR is high, the
centralized scheme offers a higher performance in comparison to the distributed
scheme. However, one should note that at realistic SNRs,
the distributed solution still achieves almost the same performance and at the
same time scales smoothly with the network size,
while a centralized solution can only be implemented in very restricted
scenarios.

\subsection{Global Optimality}
In the following, we compare the performance of the distributed algorithm to
an (nearly) optimal solution based on BNB. To simplify the analysis in this section and due to the high complexity of
the BNB algorithm, we restrict ourselves to the simple case of two sectors with two users (no mobility) and two PRBs
and a single antenna.

Given a certain tolerance $\varepsilon$, BNB converges to an optimal solution of (\ref{eq:orgproblem}).
However, it has exponential complexity and is therefore only feasible in very small settings.
We compare our distributed coordination algorithm with an equal power non-cooperative scheduler
and the BNB algorithm described in Section~\ref{sec:BNB}.

\begin{figure}[!h]
%\hrulefill
\centering
\vskip-5pt
\includegraphics[width=0.7\linewidth]{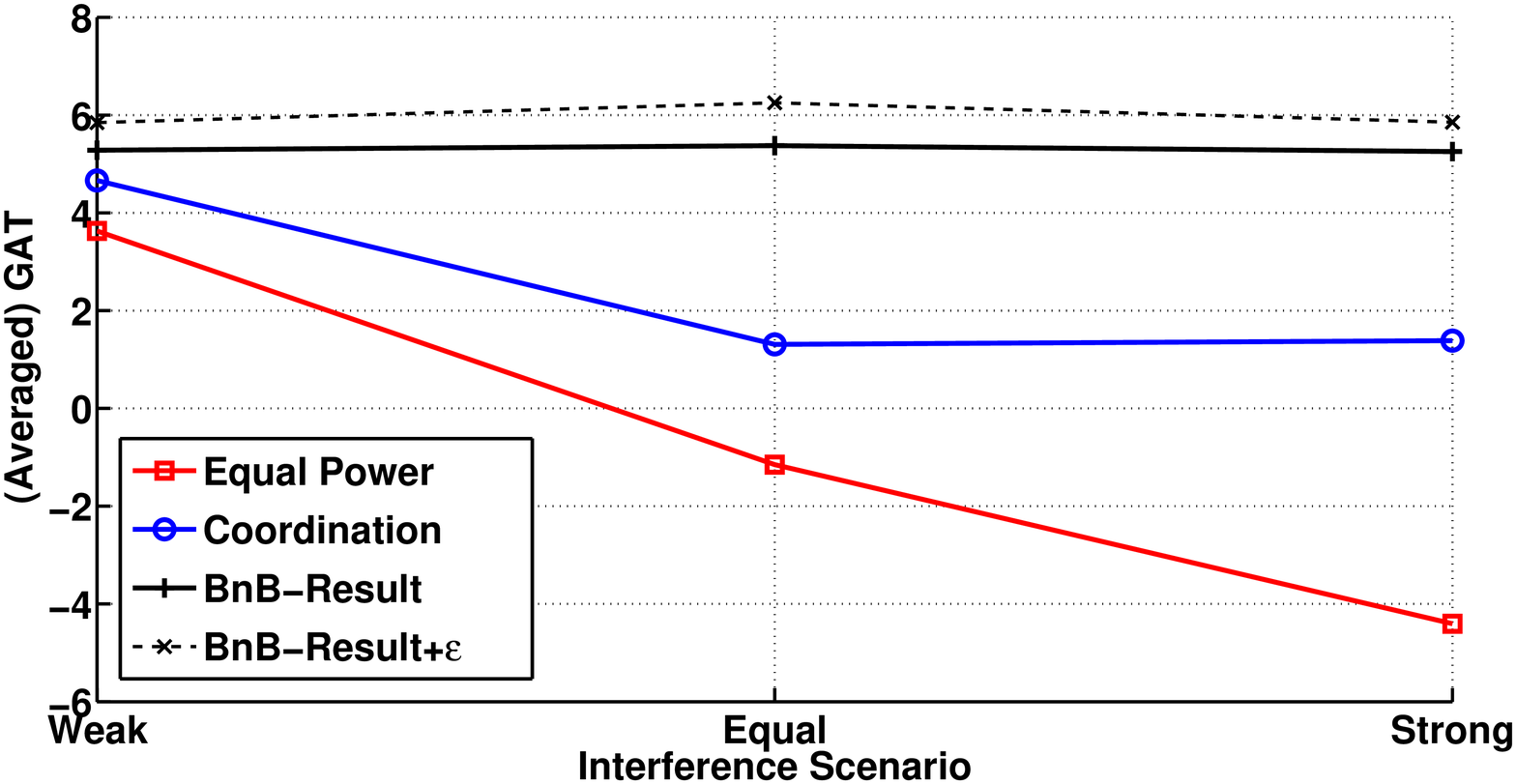}
\caption{Comparison of the distributed power control with BNB solution for different interference regimes.}
\label{fig12}
\end{figure}

\begin{figure}[!h]
\centering
\includegraphics[width=0.7\linewidth]{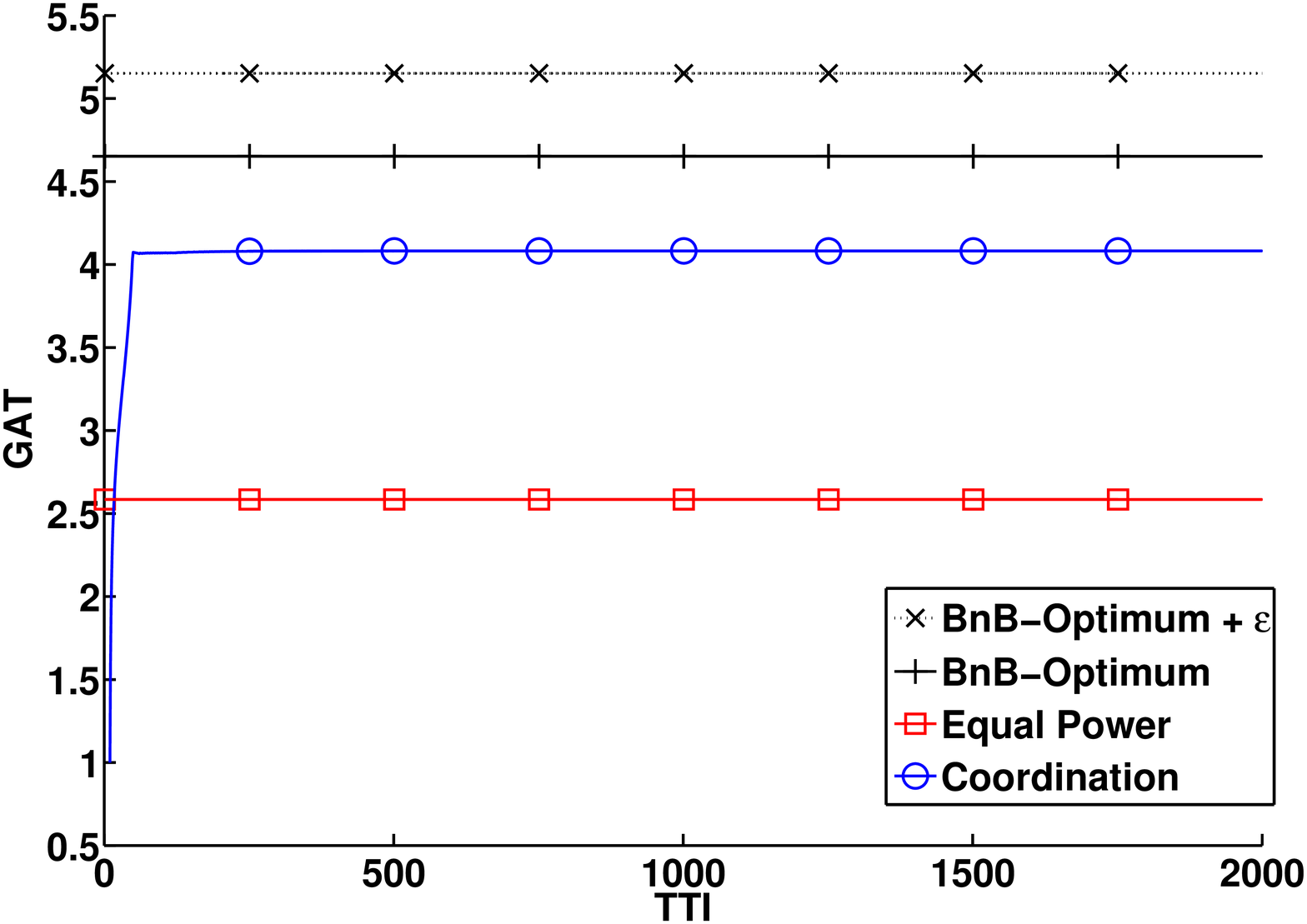}
\caption{Comparison of distributed power control with the BNB solution. Weak interference example.}
\label{fig13}
\vskip-5pt
%\hrulefill
\end{figure}

In Figures~\ref{fig12} and \ref{fig13}, the blue line represents the GAT obtained with the coordination algorithm,
while the red line represents the outcome of a non-cooperative equal-power scheduler.
The solid black line indicates the outcome of BNB. Finally, the dotted black line depicts the configured
$\varepsilon$ tolerance. Thus, the true optimum is guaranteed to lie in between the solid and dotted black lines.
To get an overview of the performance in different scenarios, we investigate three different settings, namely,
a setting with weak interference, a setting with approximately equal interference, and a setting with high interference.
In Figure~\ref{fig12}, the left marker corresponds to the weak interference case, the middle marker to the case
of equal interference, while the right marker corresponds to the case of strong interference.
The weak interference is characterized by $G_{ij}^{m} > G_{ij}^{m^\prime} (m\neq m^\prime) \quad \text{for all }
i,j,m,m^\prime$; thus, for all users, the link gains to the serving base station are higher than the gains to the
interfering base station. Note that this is, from a practical point of view, the more interesting case since due to
handover algorithms, most of the time, the gains to the own base station are stronger than to the interferers.
%\textcolor{red}{\bf\{AU Query: The previous sentence was modified. Please confirm if the intended meaning has been retained.\}}
%ANSWER: The modification is correct, the intended meaning has been retained. 
It can be observed that the distributed power control performs quite well since
we see a significant improvement over equal power scheduling. Moreover, the distributed power control appears to
realize already a large portion of the available gains.

In the `equal' interference case, the gains to the interfering base stations are roughly the same as the gains to
the own base stations; thus, $G_{ij}^{m} \approx G_{ij}^{m^\prime} (m\neq m^\prime) \;  \text{for all }
i,j,m,m^\prime$. It can be observed that the power control algorithm leaves a larger portion of the available performance
gains unused, compared to the weak interference case, although it shows significant improvements over
non-cooperative scheduling.

The strong interference case is characterized by $G_{ij}^{m} < G_{ij}^{m^\prime} (m\neq m^\prime)$, for all
$i,j,m,m^\prime$; thus, for all users, the link gains to the interfering base stations are higher
than the gains to the `own' base stations. Here, the gap in the non-cooperative algorithm's performance,
compared to the BNB result, is significantly worse than in the other cases. By contrast, the power control algorithm
shows roughly the same performance gap than in the case of equal gains.

Again, to illustrate the convergence, Figure~\ref{fig13} gives an example of the performance over time (here, for the
weak interference case). Obviously, the power control algorithm converges to a local maximum soon.

\section{Conclusions} \label{sec:Conclusions}

We proposed and compared three distributed algorithms for autonomous interference coordination in cellular SDMA networks.
The algorithms are based on a virtual layer that models the interference interdependencies in the network and
gradually adapts power control levels. The proposed algorithms differ in granularity of power control, required feedback,
signaling overhead, and the virtual model itself. System-level simulations indicate high gains both in overall utility
and in cell-edge user throughput for all three algorithms in static environments without user mobility. While VSA
offers a very fine granularity of power control on a per-beam level, it suffers from the lack of freedom to
instantaneously perform power allocation in an opportunistic manner. In fact, in an environment with significant user
mobility, only CBA achieves significant gains in both metrics. This demonstrates the superiority of CBA's approach
to enforce average power constraints but instantaneously allowing opportunistic scheduling. Comparisons to a
centralized benchmark scheme reveal that although the convergence of the centralized scheme is much faster than
in the distributed case, the performance in overall utility is comparable.

\section*{Appendices}

\section*{Appendix 1: proof of Theorem~\ref{Thm:2}}
\label{sec:Appx1}
First, we show that $f_{\varepsilon}(R)$ is everywhere differentiable. For this
purpose, we use the following Lemma~\ref{lemma: 1} and set $H(x,y) := U(R,\Phi)$.

\begin{mylemma}
\label{lemma: 1}
Consider a function $H\left(  x,y\right)$, $x\in\left[a_1,a_2\right]$, $y\in \left[a_3,a_4\right]$,
with some finite $a_1<a_2$ and $a_3<a_4$. Assume that $H\left(x,y\right)$ and its partial derivative on $x$ are
continuous, and $H\left(x,y\right)$ is concave in $x$ (for each $y$) and strictly concave in $y$ (for each $x$).
Then, the function

\begin{equation}
\varphi\left(  x\right)  =\max_{y}H\left(  x,y\right)
\label{eq:2}
\end{equation}

is continuously differentiable in $\left[a_1,a_2\right]$. (This can be generalized to the multi-dimensional case.
Namely, the domains $\left[a_1,a_2\right]$ and $\left[a_3,a_4\right]$ can be
replaced by arbitrary convex compact sets in finite-dimensional vector spaces, and derivatives replaced by gradients.)
\end{mylemma}

Note, that Lemma~\ref{lemma: 1} cannot be applied directly to (\ref{eq:VSAoptimization}) since strict concavity
in $y$ in this case is not given. Clearly, $R$ and $\Phi$ are compact since $\phi_{ijb}\in\lbrack0,1]$ and
$R_{ijb}\in\lbrack c,B],B<\infty, c>0$. Moreover, the concavity of $U$ follows by definition.
Now, we have to show that $f_{\varepsilon}(R)$ converges for any zero sequence $\varepsilon_{n}\rightarrow0$ to $f$.
First, fix a zero sequence $\varepsilon_{n}^{\ast}$. Moreover, we define a sequence of functions

\begin{align*}
f_{n}\left(  R\right)  &:=\max_{\phi_{ijb}^{m}}\sum_{i}U^m\left(  \sum_{j}
\sum_{b}\left(  \phi_{ijb}^{m}\right)  ^{1-\varepsilon_{n}^{\ast}}R_{ijb}
^{m}\right)\\  &\hat{=}f_{\varepsilon_{n}^{\ast}}(R),
\end{align*}

with $(n\in\mathbb{N})$. We can use Dini's Theorem to show that $\{f_{n}(R)\}_{n\in\mathbb{N}}$ converges uniformly to $f$.
This theorem states that if $\{f_{n}\}_{n\in\mathbb{N}}$ ($f_{n}:K\rightarrow\mathbb{R},n\in\mathbb{N}$, being a
sequence of continuous functions and $K$ being a compact metric space) converges pointwise to $f$
($f:K\rightarrow\mathbb{R}$ being a continuous function) and if $f_{n}(x)\geq f_{n+1}(x)$ ($\forall x\in K$ and $\forall
n\in\mathbb{N}$), then $\{f_{n}\}_{n\in\mathbb{N}}$ converges uniformly to $f$.

The rate space $K:=[c,B]^{IJN}$ is clearly compact, since $R_{ijb} \in\lbrack c,B],B<\infty, c>0$. Moreover,
for each $x^{\ast}\in K$, $f_{n}(x^{\ast})$ converges to $f(x^{\ast})$ when $n\rightarrow\infty$
(since $\varepsilon_{n}\rightarrow0$), and since $\phi_{ijb}\in\lbrack0,1]$, we have
$\phi^{1-\varepsilon_{n}}\geq\phi^{1-\varepsilon_{m}}$ ($\forall i,j,b$) if
$m\geq n$. Thus, $f_{n}(x)\geq f_{n+1}(x)$.

\section*{Appendix 2: proof of Lemma~\ref{lemma:pow_alloc_subprob_convexify}}
\label{sec:Appx2}
Following the approach in \cite{Papandriopoulos2009}, the convexity of optimization
problem (\ref{eq:PAsubproblemobjective} to \ref{eq:PAsubproblempowconstr}) can be
easily shown by noting that $\tilde{R}_{ijb}^m$ in (\ref{eq:PAsubproblemobjective}) can be
written as

 \[
 \alpha_{ijb}^m\log\left(\tilde{F}_{ijb}^m\right)+\beta_{ijb}^m,
 \]
where $\log(\tilde{F}_{ijb}^m)$ can be further decomposed into
\[
\log(G_{ijb}^m)+\tilde{P}_{jb}^m-\log\left(\sigma^2+\sum_{b^{\prime}\neq
b}G_{ijb^{\prime}}^{m} e^{\tilde{P}_{jb^{\prime}}^{m}}+\sum_{m^\prime \neq
m}\sum_b G_{ijb}^{m^\prime}e^{\tilde{P}_{jb}^{m^\prime}}\right).
\]
Due to the convexity of log-sum-exp \cite{boyd2004}, the term
\[
\log\left(\sigma^2+\sum_{b^{\prime}\neq b}G_{ijb^{\prime}}^{m}
e^{\tilde{P}_{jb^{\prime}}^{m}}+\sum_{m^\prime \neq m}\sum_b
G_{ijb}^{m^\prime}e^{\tilde{P}_{jb}^{m^\prime}}\right)
\]
is convex, and thus, $\tilde{R}_{ijb}^m$ is concave.
Noting that non-negative weighted addition and scalar composition preserve
concavity concludes the proof.


\begin{thebibliography}{10}

\providecommand{\url}[1]{[#1]}
\providecommand{\urlprefix}{}

%==Ref29
\bibitem{Wunder10}
G Wunder,
M Kasparick,
A Stolyar,
H Viswanathan,
{{Self-organizing distributed inter-cell beam coordination in cellular networks with best effort traffic}}.
In \textit{Proc. 8th International Symposium on Modeling and Optimization in Mobile, Ad Hoc, and Wireless Networks (WiOpt)}
(Avignon, 2010)


\bibitem{Kasparick2011}
M Kasparick,
G Wunder,
{{Autonomous Distributed Power Control Algorithms for Interference Mitigation in Multi-Antenna Cellular Networks}}
In \textit{Proc. 17th European Wireless Conference}
(Vienna, 2011), pp. 1--8


%==Ref1
\bibitem{Lee2012}
D Lee,
H Seo,
B Clerckx,
E Hardouin,
D Mazzarese,
S Nagata,
K Sayana,
{{Coordinated multipoint transmission and reception in LTE-advanced: deployment scenarios and operational challenges}}.
IEEE Commun. Mag.
\textbf{50}(2),
148--155 (2012)
%\textcolor{red}{\bf\{AU Query: Please confirm if the journal name, volume and issue numbers and page range are correct.
%Otherwise, please provide the details.\}}
%ANSWER: The data provided above is correct. 

% If possible, the following reference should be added: 
% %==Ref2
\bibitem{5gnow2014}
G. Wunder,
P. Jung,
M. Kasparick,
T. Wild,
F. Schaich,
Y. Chen,
S. ten Brink,
I. Gaspar,
N. Michailow,
A. Festag,
L. Mendes,
N. Cassiau,
D. Ktenas,
M. Dryjanski,
S. Pietrzyk,
B. Eged,
P. Vago,
F. Wiedmann,
{{5GNOW: Non-Orthogonal, Asynchronous Waveforms for Future Mobile Applications}}.
{IEEE Commun. Mag.}
\textbf{52}(2),
97--105 (2014)

%==Ref2
\bibitem{irmer2011coordinated}
R Irmer,
H Droste,
P Marsch,
M Grieger,
G Fettweis,
S Brueck,
HP Mayer,
L Thiele,
V Jungnickel,
{{Coordinated multipoint: concepts, performance, and field trial results}}.
{IEEE Commun. Mag.}
\textbf{49}(2),
102--111 (2011)

%==Ref3
\bibitem{3GPPComp2013}
3GPP, 
%\textcolor{red}{\bf\{AU Query: Please confirm if the author is correct.\}},
%ANSWER: The author is correct.
\emph{Technical Report 36.819 V11.2.0: Coordinated Multi-point Operation for LTE Physical Layer Aspects}.
(3GPP, Porto, 2013).
%\textcolor{red}{\bf\{AU Query: Please provide the place of publication to complete the details.\}}
%ANSWER: The place was added above. 

%==Ref4
\bibitem{Boudreau2009}
G Boudreau,
J Panicker,
N Guo,
R Chang,
N Wang,
S Vrzic,
{{Interference coordination and cancellation for 4G networks.}}
IEEE Commun. Mag.
\textbf{47}(4),
74--81
(2009) 
%\textcolor{red}{\bf\{AU Query: Please confirm if the completed details are correct. Please also check references
%[15, 20, and 23].\}}
%ANSWER: The completed details are correct, as well as references [15,20,23].

%==Ref5
\bibitem{Aliu2012}
O Aliu,
A Imran,
M Imran,
B Evans,
{{A survey of self organisation in future cellular networks}}.
{IEEE Commun. Surv. \& Tutor.}
\textbf{PP}(99),
1--26
(2012)

%==Ref6
\bibitem{WCM:WCM1088}
RY Chang,
Z Tao,
J Zhang,
CCJ Kuo,
{{Dynamic fractional frequency reuse (D-FFR) for multicell OFDMA networks using a graph framework}}.
{Wireless Commun. Mobile Comp.}
\textbf{13},
12--27
(2013). doi:10.1002/wcm.1088.


%==Ref7
\bibitem{Huang2009}
J Huang,
VG Subramanian,
R Agrawal,
R Berry,
{{Joint scheduling and resource allocation in uplink OFDM systems for broadband wireless access networks}}.
{IEEE J. Select. Areas Commun.}
\textbf{27}(2),
226--234 (2009)

%==Ref8
\bibitem{Venturino2009}
L Venturino,
N Prasad,
X Wang,
{{Coordinated scheduling and power allocation in downlink multicell OFDMA networks}}.
{IEEE Trans. Vehic. Technol.}
\textbf{58}(6),
2835--2848 (2009)

%==Ref9
\bibitem{gesbert2007}
D Gesbert,
SG Kiani,
A Gjendemsj,
GE Ien,
{{Adaptation, coordination, and distributed resource allocation in interference-limited wireless networks}}.
{Proc. IEEE}
\textbf{95}(12),
2393--2409 (2007)

%==Ref10
\bibitem{Yu2011}
W Yu,
T Kwon,
C Shin,
{{Multicell coordination via joint scheduling, beamforming and power spectrum adaptation}}.
In \textit{Proc. IEEE International Conference on Computer Communications (INFOCOM)}
(Shanghai, 2011),
pp. 2570--2578.
%\textcolor{red}{\bf\{AU Query: For references [10, 11, 12, 13, 14, 21, 22, 24, 25, and 29], please provide the
%editor(s), publisher, and the place of publication to complete the details.\}}
%ANSWER: Above mentioned papers (as well as reference 19) are all part of proceedings of various conferences. To clarify this we added the abbreviation ``Proc.''
% in the references. Publisher is IEEE of all the mentioned proceedings, and the place of publication (place of the conference??) was added to the 
%references. Since the references refer to conference proceedings, maybe mentioning an editor is not necessary. 

%==Ref11
\bibitem{Borst2011}
S Borst,
M Markakis,
I Saniee,
{{Distributed power allocation and user assignment in OFDMA cellular networks}}.
In \textit{Proc. 49th Annual Allerton Conference on Communication, Control, and Computing}
(Monticello, 2011), pp. 1055--1063.

%==Ref12
\bibitem{6364444}
IH Hou,
CS Chen,
{{Self-organized resource allocation in LTE systems with weighted proportional fairness}}.
In \textit{Proc. IEEE International Conference on Communications (ICC)}
(Ottawa, 2012), pp. 5348--5353.

%==Ref13
\bibitem{Infocom09}
AL Stolyar,
H Viswanathan,
{{Self-organizing dynamic fractional frequency reuse for best-effort traffic through distributed inter-cell coordination}}.
In \textit{Proc. IEEE International Conference on Computer Communications (INFOCOM)}
(Rio de Janeiro, 2009)

%==Ref14
\bibitem{Rengarajan2010}
B Rengarajan,
AL Stolyar,
H Viswanathan,
{{Self-organizing dynamic fractional frequency reuse on the uplink of OFDMA systems}}.
In \textit{Proc. 44th Annual Conference on Information Sciences and Systems (CISS)}
(Princeton, 2010), pp. 1--6.

%==Ref15
\bibitem{ETT:ETT2554}
NUL Hassan,
M Assaad,
{{Downlink beamforming and resource allocation in multicell MISO-OFDMA systems}}.
{Trans. Emer. Telecommun. Technol.}
\textbf{25}(2),
173--182 (2012)


%==Ref16
\bibitem{Papandriopoulos2009}
J Papandriopoulos,
JS Evans,
{{SCALE: a low-complexity distributed protocol for spectrum balancing in multiuser DSL networks}}.
{IEEE Trans. Inform. Theory}
\textbf{55}(8),
3711--3724 (2009)

%==Ref17
\bibitem{Papandriopoulos2008}
J Papandriopoulos,
S Dey,
J Evans,
{{Optimal and distributed protocols for cross-layer design of physical and transport layers in MANETs}}.
{IEEE/ACM Trans. Netw.}
\textbf{16}(6),
1392--1405 (2008)

%==Ref18
\bibitem{Lin2008}
B Song,
YH Lin,
RL Cruz,
{{Weighted max-min fair beamforming, power control, and scheduling for a MISO downlink}}.
{IEEE Trans. Wireless Commun.}
\textbf{7}(2),
464--469 (2008)

%==Ref19
\bibitem{Yu2012}
L Yu,
E Karipidis,
EG Larsson,
{{Coordinated scheduling and beamforming for multicell spectrum sharing networks using branch and bound}}
In \textit{Proc. 20th European Signal Processing Conference (EUSIPCO)}
(Bucharest, 2012), pp. 819--823
%\textcolor{red}{\bf\{AU Query: Please provide the volume number and page range. If this is a book,
%please provide the publisher and the place of publication.\}}
%ANSWER; Above reference is a conference paper, the missing details were added accordingly.

%==Ref20
\bibitem{Utschick2012}
W Utschick,
J Brehmer,
{{Monotonic optimization framework for coordinated beamforming in multicell networks.}}
IEEE Trans. Signal Processing
\textbf{60}(4),
1899--1909 (2012)

%==Ref21
\bibitem{Honig2008}
C Shi,
RA Berry,
ML Honig,
{{Distributed interference pricing for OFDM wireless networks with non-separable utilities}}.
In \textit{Proc. 42nd Annual Conference on Information Sciences and Systems (CISS)}
(Princeton, 2008), pp. 755--760.

%==Ref22
\bibitem{Ahmed2013}
F Ahmed,
AA Dowhuszko,
O Tirkkonen,
R Berry,
{{A distributed algorithm for network power minimization in multicarrier systems}}.
In \textit{Proc. IEEE 24th International Symposium on Personal Indoor and Mobile Radio Communications (PIMRC)}
(London, 2013), pp. 1914--1918.

%==Ref23
\bibitem{Sengupta2010}
S Sengupta,
M Chatterjee,
KA Kwiat,
{{A game theoretic framework for power control in wireless sensor networks.}}
IEEE Trans. Comput.
\textbf{59}(2),
231--242 (2010)


%==Ref24
\bibitem{Son10}
K Son,
S Lee,
Y Yi,
S Chong,
{{Practical dynamic interference management in multi-carrier multi-cell wireless networks: a reference user based approach}}.
In \textit{Proc. 8th International Symposium on Modeling and Optimization in Mobile, Ad Hoc, and Wireless Networks (WiOpt)}
(Avignon, 2010)

%==Ref25
\bibitem{6623394}
AA Dowhuszko,
F Ahmed,
O Tirkkonen,
{{Decentralized transmit beamforming scheme for interference coordination in small cell networks}}.
In \textit{Proc. First International Black Sea Conference on Communications and Networking (BlackSeaCom)}
(Batumi, 2013), pp. 121--126.

%==Ref26
\bibitem{803503}
Q Zhang,
SA Kassam,
{{Finite-state Markov model for Rayleigh fading channels}}.
{Commun. IEEE Trans.}
\textbf{47}(11),
1688--1692 (1999)

%==Ref27
\bibitem{Stolyar05}
AL Stolyar,
{{On the asymptotic optimality of the gradient scheduling algorithm for multiuser throughput allocation}}.
{Oper. Res.}
\textbf{1},
12--25 (2005)

%==Ref28
\bibitem{Stolyar2005}
A Stolyar,
{{Maximizing queueing network utility subject to stability: greedy primal-dual algorithm}}.
{Queueing Syst.}
\textbf{50}(4),
401--457 (2005)

%==Ref30
\bibitem{boyd2004}
S Boyd,
L Vandenberghe,
\textit{{Convex Optimization}}
(Cambridge University Press, Cambridge, 2004)

%==Ref31
\bibitem{6129545}
E Matskani,
ND Sidiropoulos,
L Tassiulas,
{{Convex approximation algorithms for back-pressure power control}}.
{IEEE Tran. Signal Process.}
\textbf{60}(4),
1957--1970 (2012)

%==Ref32
\bibitem{marks1978}
BR Marks,
GP Wright,
{{Technical note-a general inner approximation algorithm for nonconvex mathematical programs}}.
{Oper. Res.}
\textbf{26}(4),
681--683 (1978)

%==Ref33
\bibitem{GlobOpt}
R Horst,
PM Pardalos,
NV Thoai,
\textit{{Introduction to Global Optimization}}, 2nd edn.
(Kluwer Academic Publishers, Dordrecht, 2000)
%\textcolor{red}{\bf\{AU Query: Please provide the place of publication to complete the details.\}}
%ANSWER: The place of the publisher of this book was added above. 

%==Ref34
\bibitem{BeamDis2007}
{3GPP R1-073937, Alcatel-Lucent},
\textit{{3GPP TSG RAN WG1 \#50bis: Comparison Aspects of Fixed and Adaptive Beamforming for LTE Downlink}}.
(3GPP, Shanghai, 2007)

%==Ref35
\bibitem{WINNER5.4}
Wireless World Initiative New Radio, % \textcolor{red}{\bf\{AU Query: Please confirm if the author is correct.\}},
%ANSWER: The author is correct. 
\emph{D5.4 Final Report on Link and System Level Channel Models, Technical report, IST-2003-507581}
(WINNER, 2005)
%\textcolor{red}{\bf\{AU Query: Please provide the place of publication to complete the details.\}}
%ANSWER: This is a European funded project deliverable. I am not sure what the place of publication could be.
%It is available at: http://www.ist-winner.org/DeliverableDocuments/D5.4.pdf

\end{thebibliography}
\end{document}